\documentclass[prb,twocolumn,showpacs,superscriptaddress]{revtex4-1}

\usepackage{graphicx}
\usepackage{amsmath,amssymb,bm}

\usepackage{color}

\begin{document}


\title{
Exact Green's function for a multi-orbital Anderson impurity 
at  high bias voltages
}


\author{Akira Oguri}
\affiliation{
Department of Physics, Osaka City University, Sumiyoshi-ku, 
Osaka 558-8585, Japan
}

\author{Rui Sakano}
\affiliation{
The institute for Solid State Physics, The University of Tokyo, Kashiwa, Chiba 277-8581, Japan
}

\date{\today}


\begin{abstract}
We study the nonequilibrium Keldysh Green's function 
 for an $N$-orbital Anderson model at high bias voltages, 
extending  a previous work  
which  for the case only with the spin degrees of freedom $N=2$, 
to arbitrary $N$. 
Our approach uses an effective non-Hermitian Hamiltonian that is  
defined with respect to  a Liouville-Fock space  
in the context of a thermal field theory.
The result correctly  captures the relaxation processes at high energies, 
and is asymptotically 
exact not only in the high-bias limit 
but also in the high-temperature limit at thermal equilibrium. 
We also present an explicit continued-fraction representation 
of the Green's function. It clearly shows that  
the imaginary part is recursively determined by 
the decay rate of intermediate states 
with at most $N-1$ particle-hole-pair excitations.
These high-bias properties 
follow from the conservations of a generalized charge and current 
in the Liouville-Fock space.
We also examine temperature dependence of
the spectral function in equilibrium, comparing the exact results 
with the  numerical finite-$T$ and analytical $T \to \infty$ 
results of the non-crossing approximation (NCA).

\end{abstract}

\pacs{72.15.Qm, 73.63.Kv, 75.20.Hr}

\maketitle

\section{Introduction}
\label{sec:introduction}

Role of the orbital degrees of freedom has been one of the key issues 
in quantum dots and dilute magnetic alloys.\cite{Hewson_book,Yoshimori}
It also gives a variety in the many-body effects,
such as the Kondo effect and Coulomb blockade, 
 in a wide energy scale.
\cite{Grobis,ScottNatelson,Heiblum,Kobayashi,Delattret,Basset}
Specifically, 
the orbital degeneracy affects the nonequilibrium 
current and current noise 
of quantum dots driven by the bias voltage $eV$, and has recently 
been studied in the low-energy 
Fermi-liquid regime.\cite{Mora2009,Sakano,SakanoFCS,SakanoHund} 

However, further investigations of higher energy 
regions beyond the Fermi-liquid regime are still needed 
to comprehensively explore the orbital effects 
 on the correlated electrons in quantum dots.  
There are some efficient numerical approaches 
that can provide information relevant 
to the intermediate energy regions. 
For instance, the Wilson numerical renormalization group (NRG),\cite{Anders} 
the density matrix renormalization group,\cite{Kirino} 
the continuous-time 
quantum Monte Carlo methods\cite{Werner,MuhlbacherUrbanKomnik}
and the Matsubara-voltage approach,\cite{Han}  
can be applied to the multi-orbital Anderson model for quantum dots 
in the case where the internal degrees of freedom $N$ are not so large.
Alternatively, perturbative large $N$ approaches, 
such as the non-crossing approximation (NCA) \cite{Bickers,KeiterQin,WM,Kroha,OtsukiKuramoto} 
and  $1/(N-1)$ expansion\cite{ao_N_symmetric,ao_N_away} 
can explore the parameter regions complementary 
to the numerical ones.

We have previously considered the high-bias limit of 
the $N=2$ Anderson model,\cite{AO2002,OguriSakano}
for which 
nonequilibrium quantum transport in the low-energy Fermi-liquid region 
 has been investigated for a long time. 
\cite{HDW,MW,KNG,ao2001,HBA,GogolinKomnik,Sela2006,Golub,Fujii2010}
We have shown that the Keldysh Green's function\cite{Keldysh,Caroli} 
 is solvable in the opposite limit $eV \to \infty$, 
 where the excitations of whole energy scales 
equally contribute to the dynamics. 
In this limit, the model can be mapped onto 
a non-Hermitian Hamiltonian of two effective sites 
in a doubled Hilbert space that is defined  
in the thermal field theory.\cite{UmeMatTac,EzaAriHas} 
The asymptotically exact Green's function for $eV \to \infty$ 
has a similar form to the atomic-limit solution 
of Hubbard I,\cite{HubbardI,Donianch,HaugJauho} 
but is still non-trivial as 
the hybridization energy scale $\Delta$ ($\equiv \Gamma_L+\Gamma_R$)
that competes with the Coulomb repulsion $U$ 
is fully taken into account without any assumptions. 
For this reason, 
the result correctly captures the imaginary part 
due to the relaxation processes, which in the high-bias limit    
is determined by the damping of 
 a single particle accompanied by a virtually excited 
particle-hole pair in the intermediate states.
Furthermore, it has also been clarified that 
the spectral weight depends sensitively on the asymmetry 
 in $\Gamma_L$  and  $\Gamma_R$, which are 
the hybridizations between the impurity 
and the reservoirs on the left and right, respectively.

In the present paper,  we extend the formulation to treat  
the multi-orbital Anderson model,
and provide the asymptotically exact high-bias Green's function 
for generic two-body interactions $U_{mm'}$ between 
the electrons in different orbitals $m$ and $m'$ 
with orbital-dependent hybridizations
 $\Gamma_{L,m}$ and  $\Gamma_{R,m}$.
The thermal-field-theoretical approach\cite{UmeMatTac,EzaAriHas} 
that we use is equivalent to the Keldysh formalism.
 However, the time evolution along 
the backward Keldysh contour is dealt with 
in a different way, 
using fictitious fermions    
defined with respect to the enlarged Hilbert space.  
It is also referred to as a Liouville-Fock space   
and has been applied to quantum-transport problems.
\cite{EspositeHarolaMukamel,DzioevKosov,SaptosovWegewijs,SaptosovWegewijs2014}
We show that   
the effective non-Hermitian Hamiltonian 
can be expressed in terms of a generalized charge 
and current, 
which commute each other,  also in the multi-orbital case.
This algebraic structure makes the many-body effects on the 
Green's function and other dynamic correlation functions 
separable in the time representation.
The exact Green's function 
can be expressed in a factorized form,   
which consists of contributions 
of the intermediate particle-hole pair excitations 
from each of the orbitals.

For the $m$-independent interactions and hybridizations, 
namely for $U$, $\Gamma_{L}$ and $\Gamma_{R}$, 
we also obtain the continued fraction representation 
of the  Green's function for arbitrary $N$ 
as a function of  frequency $\omega$. 
The results show that the spectral function 
has $N$ distinguishable peaks,  
the height of which is determined by 
the binominal distribution, 
specifically  for symmetric hybridizations $\Gamma_L = \Gamma_R$ 
and strong interactions $U \gg \Delta$. 
In the continued fraction representation, 
the imaginary part due to the relaxation 
of the intermediate state with $k$ ($=1,2,\ldots,N-1$) particle-hole 
pair excitations  
recursively emerges 
through the iteration that terminates after $N-1$ steps.

Our results also describe the high-temperature 
limit at equilibrium $eV=0$, 
where the Fermi function becomes an $\omega$ 
independent constant, $f(\omega)\to 1/2$.  
We also examine temperature dependence of 
the spectral function of the particle-hole 
symmetric SU(4) Anderson model, 
using the NCA 
which can also be analytically solved in the limit of $T\to \infty$.
Near the Kondo temperature $T \simeq T_K$,  
besides the Kondo peak at the Fermi level $\omega = 0$, 
not all the four sub-peaks of the atomic nature 
can be seen yet but the lower two sub-peaks 
can be at $\omega = \pm U/2$. 
It is at much higher temperatures $T\gg T_K$ that 
the higher-energy sub-peaks emerge at $\omega = \pm 3U/2$. 
The NCA reasonably describes these features of the temperature dependence  
although there are some quantitative deviations 
from the exact results in the high-temperature limit. 
The analytic solution can also be used in such a way 
as a standard for comparisons to check the accuracy of 
any approximations.

This paper is organized as follows. 
We describe the relation between the  Keldysh formalism 
and the thermal-field-theoretical approach 
in Sec.\ \ref{sec:Keldysh_formalism}.
The effective non-Hermitian Hamiltonian 
for the high-bias limit is introduced 
 in Sec.\  \ref{sec:effective_action}. 
The initial and final states 
for the time-dependent perturbation theory 
are introduced with the nonequilibrium density matrix  
for the Liouville-Fock space in Sec.\ \ref{sec:boundary_condition_TFD}. 
General properties of the dynamic correlation functions,  
which can be deduced from the charge and current conservations 
in the Liouville-Fock space for $eV \to \infty$,  
are discussed in Sec.\ \ref{sec:correlation functions}.
The derivation of exact high-bias Green's  function 
for generic two-body interactions $U_{mm'}$ 
is given in Sec.\ \ref{sec:calculations_for_G}. 
The  continued fraction representation 
of the Green's function 
for the $m$ independent interaction $U$ and 
properties of the spectral function 
are described in Sec.\ \ref{sec:uniform_interaction}.
Summary is given in Sec.\ \ref{sec:summary}.


\section{Keldysh formalism}
\label{sec:Keldysh_formalism}

We start with a multi-orbital Anderson impurity 
coupled to two noninteracting leads 
($\alpha =L,\,R$). The Hamiltonian is given by   
${\cal H} =   {\cal H}_0 + {\cal H}_U$ with   
\begin{align}
{\cal H}_0 =& \   
 \sum_{m=1}^N 
\varepsilon_{d,m}^{} \,n_{d,m}^{} 
+ \sum_{\alpha=L,R}\,
\sum_{m=1}^N 
 v_{\alpha,m}^{} \left(
d_{m}^{\dagger} 
\psi^{}_{\alpha m} 
+ \mbox{H.c.} \right) 
 \nonumber \\
& \   + 
\sum_{\alpha=L,R}
\sum_{m=1}^N 
\int_{-D}^D  \! d\epsilon\,  \epsilon\, 
 c^{\dagger}_{\epsilon \alpha m} c_{\epsilon \alpha m}^{} \;, 
\label{Hami_seri_part} 
\\
{\cal H}_U 
=& \     \frac{1}{2} \sum_{m \neq m'} U_{mm'}^{} \, 
n_{d,m}^{} n_{d,m'}^{} \;. 
\label{eq:mia-model} 
\end{align}
Here, $n_{d,m}^{} = d_{m}^{\dagger} d_{m}^{}$ 
describes the local charge in the quantum dot, 
and  $d_m^{\dagger}$ creates an electron 
in a one-particle state with 
a quantum number $m$ ($=1,2, \cdots, N$)  
whose eigenenergy $\varepsilon_{d,m}^{}$ 
generally depends on $m$, for instance, in a finite magnetic field.  
 The inter-electron   
interaction  $U_{mm'}^{}$ generally depends on $m$ and $m'$,  
with a requirement  $U_{mm'}^{}=U_{m'm}^{}$. 
The operator $c_{\epsilon\alpha m}^{\dagger}$ creates 
a conduction electron 
with energy $\epsilon$ in orbital $m$ 
for the lead  
on the left $\alpha=L$ or right $\alpha=R$. 
It is normalized such that 
$\{ c^{\phantom{\dagger}}_{\epsilon\alpha m}, 
c^{\dagger}_{\epsilon'\alpha'm'}
\} = \delta_{\alpha\alpha'} \,\delta_{mm'}   
\delta(\epsilon-\epsilon')$. The linear combination of 
the conduction electrons, defined by 
$\psi^{}_{\alpha m} \equiv  \int_{-D}^D d\epsilon \sqrt{\rho_c^{}} 
\, c^{\phantom{\dagger}}_{\epsilon\alpha m}$ 
with $\rho_c^{}=1/(2D)$, couples to the quantized levels of the dot 
via the hybridization matrix element $v_{\alpha,m}^{}$.
This hybridization causes an 
$m$-dependent level broadening of the energy scale   
 $\Delta_m \equiv \Gamma_{L,m} + \Gamma_{R,m}$ with $\Gamma_{\alpha,m} = 
\pi \rho_c^{}\, v_{\alpha,m}^2$. 
We consider the parameter region where 
the half band-width $D$  
 is much grater than the other energy scales, 
$D \gg \max( U_{mm'}, \Delta_m, |\varepsilon_{d,m}^{}|, T, eV)$ 
unless otherwise noted.

\begin{figure}[b]
 \leavevmode
\begin{minipage}{0.9\linewidth}
 \includegraphics[width=1\linewidth]{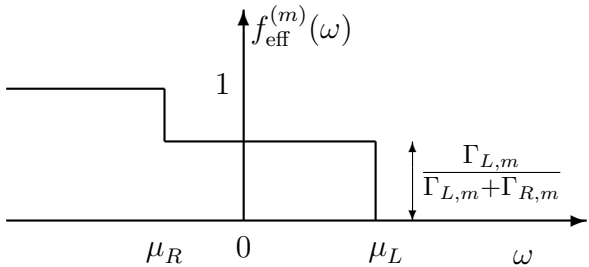}
\end{minipage}
%
%
%
%
%
%
\caption{
The nonequilibrium distribution function $f_\mathrm{eff}^{(m)}(\omega)$ 
for $\mu_L-\mu_R = eV$ and $T=0$.
The Fermi level at equilibrium, $eV=0$,
is chosen to be the origin of energy  $\omega=0$. 
}
\label{fig:distribution}
\end{figure}

Nonequilibrium steady state under a finite bias voltage  
can be described by the Keldysh formalism.\cite{Keldysh,Caroli,MW,HDW}
Specifically, we use an effective action 
$\mathcal{S} = \mathcal{S}_0+\mathcal{S}_U$ 
that determines the time evolution along the Keldysh contour,
\begin{align}
   \mathcal{Z}  \,=&  
  \int 
  \! D\overline{\eta}\, D\eta \ 
   e^{i \,\left[\,\mathcal{S}_0(\overline{\eta},\,\eta) \,+\,  
  \mathcal{S}_U(\overline{\eta},\,\eta)\,\right]}
  , 
\\
\mathcal{S}_0 =& \ 
   \sum_{m=1}^N \int_{-\infty}^{\infty} dt\,dt'\; 
 \overline{\bm{\eta}}_{m}^{}(t)\, 
        \bm{K}_{0,m}(t, t')\, 
 \bm{\eta}_{m}(t') \;, 
\label{eq:S0_Kldysh}
 \\
\mathcal{S}_U =&
   - 
\frac{1}{2} \sum_{m \neq m'} U_{mm'}^{}
\! 
\int_{-\infty}^{\infty} \!\! dt 
\,
\nonumber \\ 
& \quad    \times
 \Bigl\{\,
  \overline{\eta}_{-,m}(t)\,
  \eta_{-,m}(t)\,
  \overline{\eta}_{-,m'}(t)\,
\eta_{-,m'}(t) 
\nonumber \\ 
 & \qquad \ \   
 -    
  \overline{\eta}_{+,m}(t)\,
  \eta_{+,m}(t)\,
  \overline{\eta}_{+,m'}(t)\,
  \eta_{+,m'}(t) \Bigr\}.
\label{eq:SU_Kldysh}
\end{align}
Here, 
 $
 \overline{\bm{\eta}}_{m}
 = \left(\, 
 \overline{\eta}_{-,m}, \, 
 \overline{\eta}_{+,m} \,\right) 
 $ is a pair of the Grassmann numbers for the $-$ and $+$ 
 branches of the Keldysh contour. 
The kernel  $\bm{K}_{0,m}(t,t')$ 
is given by the Fourier transform of the noninteracting Green's function,
\begin{align}
& 
\bm{K}_{0,m}(t,t')  \, =   
      \int_{-\infty}^{\infty} \!{ d\omega \over 2 \pi }\,
      \left\{\bm{G}_{0,m}(\omega)\right\}^{-1}
      e^{-i\omega (t-t')}  \;, 
 \label{eq:K0_Keldysh} \\ 
&
\!\!\!
\left\{\bm{G}_{0,m}(\omega)\right\}^{-1} 
=  \,  (\omega-\varepsilon_{d,m}) 
\, \bm{\tau}_3^{} - \bm{\Sigma}_{0,m}(\omega) \;, 
\label{eq:G0_keldysh}
\\
& 
  \bm{\Sigma}_{0,m}(\omega) \, = \,   
-i\Delta_m \bigl[\,1-2f_\mathrm{eff}^{(m)}(\omega) \,\bigr] 
\bigl( \bm{1} -  \bm{\tau}_1^{} \bigr)
+ \Delta_m \bm{\tau}_2^{} .
\label{eq:U0_self_keldysh}
\end{align}
Here,   $\bm{1}$ is the $2\times 2$ unit matrix and
  $\bm{\tau}_j$ for $j=1,2,3$ are the Pauli matrices,  
 \begin{align}
 \!\!\!
 \bm{\tau}_1 =
 \left(
 \begin{matrix}
 0 &   1 \cr
 1 &  0 \cr  
 \end{matrix}
 \right) , 
 \quad
 \bm{\tau}_2 =
 \left(
 \begin{matrix}
 0 &   -i \cr
 i &  \ 0  \cr  
 \end{matrix}
 \right) , 
 \quad
 \bm{\tau}_3 =
 \left(
 \begin{matrix}
 1 &  \  0 \cr
 0 &  -1 \cr  
 \end{matrix}
 \right) .
 \end{align}
The distribution function 
 $f_{\mathrm{eff}}^{(m)}(\omega)$ is defined by 
\begin{equation}
f_\mathrm{eff}^{(m)}(\omega) 
\, = \, 
\frac{\Gamma_{L,m}\,f_L(\omega) + \Gamma_{R,m}\,f_R(\omega)}
 { \Gamma_{L,m} +\Gamma_{R,m} } \;,
\label{eq:f_eff}
\end{equation}
where  $f_{\alpha}(\omega) 
= [\,e^{(\omega-\mu_\alpha)/T}+1\,]^{-1}$ and
 $\mu_\alpha$ is the chemical potential for lead $\alpha$.
This distribution function 
describes the energy window as depicted in Fig.\ \ref{fig:distribution},  
and determines the long time behavior of $\bm{K}_{0,m}(t,t')$ as 
a function of $t-t'$. 
Furthermore, temperature $T$ and 
bias voltage $eV \equiv \mu_L - \mu_R$ 
 enter through  $f_\mathrm{eff}^{(m)}(\omega)$ 
for impurity correlation functions.

\section{Liouville-Fock space for $eV \to \infty$}
\label{sec:effective_action}

We consider two kinds of the high-energy limits in the present work.
One is the high-bias limit $eV\gg T$, 
where $f_L \to 1$ and $f_R \to 0$. 
The other is high-temperature limit $T \gg eV$ 
where $f_L=f_R \to 1/2$, and this 
includes thermal equilibrium at $eV=0$ as a special case.
In both of these two limits, 
the distribution function $f_\mathrm{eff}^{(m)}(\omega)$ becomes 
a constant independent of the frequency $\omega$,
\begin{align}
\!\!\!\!\!\!\!\!
f_\mathrm{eff}^{(m)}(\omega) 
\, \to \left\{  
\begin{array}{lr}
 \frac{\Gamma_{L,m}}{\Gamma_{L,m}+\Gamma_{R,m}} \;, &   \mbox{for} \ eV \to \infty 
\\
\quad \frac{1}{2}  \quad \ , & \mbox{for} \ \ \,  T \to \infty  
\rule{0cm}{0.4cm}
\end{array}
\right. \; .
\label{eq:f_eff_high_energy}
\end{align}
Then the hybridizations self-energy  $\bm{\Sigma}_{0,m}(\omega)$ 
also becomes independent of $\omega$, and  then
excitations of whole energy scales equally contribute 
to the dynamics.
This makes the problems in the high-energy limits solvable. 
In the following, we concentrate on the $eV \to \infty$ limit 
because the $T \to \infty$ limit is equivalent to 
the symmetric coupling case $\Gamma_{L,m} =\Gamma_{R,m}$ of 
the high-bias limit 
as long as local properties nears the impurity site are concerned.

\subsection{Effective non-Hermitian Hamiltonian} 

In the high-bias limit,
the hybridization self-energy 
defined in Eq.\ \eqref{eq:U0_self_keldysh}  
is given by an $\omega$ independent matrix,  
\begin{align}
& 
\lim_{eV \to \infty} \!  \bm{\Sigma}_{0,m}(\omega) 
\,=\, 
\bm{\tau}_3 \,\bm{L}_{0,m}^{} \;, 
\label{eq:L0_matrix_Vinf_tau3}
\\
& \!\!\!
\bm{L}_{0,m}^{} 
 \equiv  \, 
i\left [ \,
\begin{matrix} 
 \Gamma_{L,m} - \Gamma_{R,m}   & -2\Gamma_{L,m}\,  \cr
 -2\Gamma_{R,m}  &  -(\Gamma_{L,m} - \Gamma_{R,m})      
\rule{0cm}{0.5cm}  
\cr
\end{matrix}          
\right]  . 
\label{eq:L0_matrix_Vinf}
\end{align}
Then, the kernel $\bm{K}_{0,m}(t,t')$ takes a Markovian form 
 with a linear combination of $\delta(t-t')$ and its derivative. 
The derivative  
arises from the $\omega$ 
linear part of $\left\{\bm{G}_{0,m}(\omega)\right\}^{-1}$, 
and the noninteracting part of the action $\mathcal{S}_0$ can be expressed 
in  a single integration with respect to $t$,  
\begin{align}
& 
\! 
   \mathcal{S}_0 \to 
   \sum_{m=1}^N \! \int_{-\infty}^{\infty} \!\! dt \,  
\bm{\eta}_{m}^{\dagger}(t) 
\left\{
\bm{1}  
\!\left(\! i \frac{\partial}{\partial t}  - \varepsilon_{d,m} \!\right)\! 
- \bm{L}_{0,m}^{}   
\right\} 
 \bm{\eta}_{m}(t) . 
  \label{eq:S0_Vinf} 
\end{align}
Here, the transformation 
$\bm{\eta}_{m}^{\dagger} = \overline{\bm \eta}_{m}\,\bm{\tau}_3$  
has been introduced only for the conjugate part of the Grassmann numbers, 
 keeping the counter part ${\bm \eta}_{m}$  unchanged.  
This transform makes the time-derivative term $\mathcal{S}_0$ diagonal, 
keeping the interacting action $\mathcal{S}_U$ in a similar form 
\begin{align}
& 
\!\!\!
\mathcal{S}_U = 
   - 
\frac{1}{2} \sum_{m \neq m'} U_{mm'}^{}
\! \int_{-\infty}^{\infty} \!\! dt  \,
 \nonumber \\ 
  & \qquad   \quad 
\times
 \Bigl\{\,
  \eta_{-,m}^{\dagger}(t)\,
  \eta_{-,m}(t)\,
  \eta_{-,m'}^{\dagger}(t)\,
  \eta_{-,m'}(t) 
 \nonumber \\ 
  & \qquad   \qquad \quad 
-     
  \eta_{+,m}^{\dagger}(t)\,
  \eta_{+,m}(t)\,
  \eta_{+,m'}^{\dagger}(t)\,
  \eta_{+,m'}(t) \Bigr\}.
\label{eq:SU_TFD}
\end{align}

Therefore, in the high-bias limit, the Lagrangian that corresponds to 
the integrand of $\mathcal{S}_0 + \mathcal{S}_U$  
does not have an explicit time-dependence 
other than the first derivative $i\partial/\partial t$ 
term in Eq.\ \eqref{eq:S0_Vinf}.
The contributions of the conduction electrons 
enter through   $\bm{L}_{0,m}^{}$. 
The Lagrangian of this form can also be constructed from 
a non-Hermitian Hamiltonian  
 defined with respect to  the doubled Hilbert space, 
 consisting only of the impurity degrees of freedom:      
 $\,\widehat{H}_\mathrm{eff}^{} = \widehat{H}_\mathrm{eff}^{(0)} 
 + \widehat{H}_\mathrm{eff}^{(U)}$,  
%
\begin{align}
 \widehat{H}_\mathrm{eff}^{(0)} \equiv  & \ 
\sum_{m=1}^N 
\xi_{d,m} 
  \left( n_{-,m}^{} \! +  n_{+,m}^{}  -1 \right) 
\nonumber \\
&  + 
\sum_{m=1}^N 
\left( 
\bm{d}_{m}^{\dagger} \bm{L}_{0,m}^{} \bm{d}_{m}^{} 
 - i\Delta_m 
\right) , 
\label{eq:H_0_Vinf}
\\
\nonumber \\
 \widehat{H}_\mathrm{eff}^{(U)}   \equiv &   
\ \frac{1}{2} \sum_{m \neq m'} U_{mm'}^{}   
\Biggl[\,
  \Bigl(\!n_{-,m }^{} \! - \! \frac{1}{2}\Bigr) 
\!
  \Bigl(\!n_{-,m' }^{} \! - \!  \frac{1}{2}\Bigr) 
 \nonumber \\  %
 & \qquad \qquad \qquad \ \  
 \! - \!    \Bigl(\!n_{+,m }^{} \! - \!  \frac{1}{2}\Bigr)
\!  
  \Bigl(\!n_{+,m' }^{} \! - \!  \frac{1}{2}\Bigr)  
 \,\Biggr] .
\label{eq:H_U_Vinf}
\end{align}
Here,
\begin{align}
 \xi_{d,m} = \varepsilon_{d,m} + \frac{1}{2} \sum_{m'(\neq m)}U_{mm'}^{} \;, 
\label{eq:Xi}
\end{align}
and 
$
 {\bm{d}}_{m}^{\dagger}
 = 
\bigl( d_{-,m}^{\dagger}\,, \,  d_{+,m}^{\dagger}\bigr) 
$ is a set of two independent fermion operators 
introduced for the  $-$ and $+$ branches, respectively, 
and  $n_{\mu,m} =  d_{\mu,m}^{\dagger} d_{\mu,m}^{}$.   
In this representation, 
the fermion operators with the label \lq\lq $-$" describe the original 
impurity electron $d_{-,m}^{\dagger} \equiv d_{m}^{\dagger}$. 
The other component with the label \lq\lq $+$" corresponds to 
a tilde-conjugate operator $\widetilde{d}_{m}^{\dagger}$ 
in the standard notation 
of the thermal field theory.\cite{UmeMatTac,EzaAriHas} 
Specifically, 
our representation uses 
a particle-hole transformed version where    
 $\,d_{+,m}^{\dagger} \equiv \widetilde{d}_{m}^{}$. 

Note that the conduction degrees of freedom 
have been effectively decoupled, and 
the extended Hilbert space for the impurity states, 
which is referred to as Liouville-Fock space
\cite{EspositeHarolaMukamel,DzioevKosov,SaptosovWegewijs,SaptosovWegewijs2014}
 in the following, 
consists of $2^{2N}$ basis sets.
The time evolution of the state vectors in this space   
is described by the Heisenberg operators,\cite{UmeMatTac,EzaAriHas}  
\begin{align}
\mathcal{O}(t) \,\equiv& \  
e^{i\widehat{H}_{\mathrm{eff}} t}
\,\mathcal{O}\,
e^{-i\widehat{H}_{\mathrm{eff}} t} \;,
\\
i \,\frac{\partial \mathcal{O}(t)}{\partial t} \,= & \  
\left[\,\mathcal{O}(t)\,,\,\widehat{H}_{\mathrm{eff}} \,\right] 
\;.
\end{align}

\subsection{Charge and current representation} 
\label{subseec:current_algebra}

One of the merits of the effective 
Hamiltonian formulation is that 
it can clearly extract the properties 
that system acquires in the high-bias limit. 
In order to see the precise features, we rewrite the 
interaction part defined in Eq.\ \eqref{eq:H_U_Vinf} in the form,
\begin{align}
 \widehat{H}_\mathrm{eff}^{(U)}   = & \  
\frac{1}{2} \sum_{m \neq m'} U_{mm'}^{}   
\left( 
\,Q_{m}^{} q_{m'}^{} 
+ Q_{m'}^{} q_{m}^{} 
\right) 
\nonumber \\
 =  & \ 
\sum_{m=1}^N q_{m}^{} \, \widehat{(UQ)}_m^{}  
\;.
\end{align}
Here, the operators 
 $Q_{m}^{}$, $q_{m}^{}$, and $\widehat{(UQ)}_m^{}$ are defined by 
\begin{align}
& 
Q_{m}^{} \equiv     n_{-,m} + n_{+,m} -1,  
\qquad    
q_{m}^{} \equiv \frac{n_{-,m} - n_{+,m}}{2} ,   
\label{eq:charge_q_Q} \\
& \widehat{(UQ)}_m^{} \,  \equiv \,  
 \sum_{m' (\neq m)}  \!\! U_{mm'}^{}   Q_{m'}^{} .
\end{align}
The operator $\widehat{(UQ)}_m^{}$ 
corresponds to the potential that is induced 
in the orbital $m$ by the particles occupying  
the other orbitals $m' (\neq m)$. 
Note that this potential 
 $\widehat{(UQ)}_m$  vanishes identically 
in the subspace where $Q_{m'}=0$ for all $m'$ ($=1,2,\ldots,N$).  
This happens for the final and initial states,
$\langle \! \langle I |\! |$ 
and  
$|\! | \rho \rangle \! \rangle$,  
which are introduced in the next section 
for the time-dependent perturbation 
theory in the Liouville-Fock space.

The off-diagonal components of $\widehat{H}_{\mathrm{eff}}$  
can be regarded as the operators, equivalent to 
the current  $I_{R,m}^{}$ flowing from the dot to the right lead 
and $I_{L,m}^{}$  flowing from the left lead to the dot, 
\begin{align}
I_{R,m}^{} =  
 -2 \Gamma_{R,m} \,d_{+,m}^{\dagger}d_{-,m}^{} ,
  \quad 
 I_{L,m}^{} =  
 -2 \Gamma_{L,m} \,d_{-,m}^{\dagger}d_{+,m}^{}  .
 \label{eq:current_eV_inf_def} 
\end{align}
Although these are non-Hermitian,  
the  operator equivalence holds with respect to the Liouville-Fock space.  
Using these charge and current operators, 
the effective Hamiltonian can be expressed in the form 
\begin{align}
\widehat{H}_\mathrm{eff}^{}
 \,= & \    
\sum_{m=1}^N \xi_{d,m} \,Q_{m}^{} 
+i \sum_{m=1}^N \left( P_{m}^{} -\Delta_m \right) , 
\label{eq:H_Vinf_with_currents}
\\
 P_{m}^{} \,\equiv & \     
 I_{R,m} + I_{L,m} 
 + 
 2\, \mathcal{W}_{m}^{}\, q_{m}^{} , 
\label{eq:current_P}
 \\
 \mathcal{W}_{m}^{} \, \equiv& \        
  (\Gamma_{L,m} -\Gamma_{R,m}) - i\, \frac{1}{2}  \,\widehat{(UQ)}_m^{} \;.
\label{eq:W_overline} 
\end{align}
The two operators $Q_{m}$ and $P_{m}$  
commute each other, 
and  also commute  respectively  
with $\widehat{H}_\mathrm{eff}^{}$,  
\begin{align}
\bigl[Q_m\,,\, P_{m'} \bigr]  =0, 
\quad \ 
\bigl[Q_m\,,\, \widehat{H}_\mathrm{eff}^{} \bigr]  =0 , 
\quad  
\bigl[P_m\,,\, \widehat{H}_\mathrm{eff}^{} \bigr]  =0 . 
\label{eq:EOM_Qtot_Vinf}
\end{align}
Therefore, 
 $Q_{m}$ and $P_{m}$  are conserved, 
and  $\widehat{H}_\mathrm{eff}^{}$ acquires 
a highly symmetrical algebraic structure.
The equations of motion for the relative charge $q_{m}^{}$ 
and the relative current $p_{m}^{}$ constitute a closed system
\begin{align}
 &\frac{\partial q_{m}^{}}{\partial t} \,= \,  - \, p_{m}^{} \,, 
\qquad \qquad 
p_{m}^{} \,\equiv \,   I_{R,m} - I_{L,m} \,,   
\\
&\frac{\partial p_{m}^{}}{\partial t} 
\, =  \,  
 4\, \mathcal{L}_{m}^{2} \,q_{m}^{}
+ 
 2\,\mathcal{W}_{m}^{} P_{m}^{} \;,  
\label{eq:EOM_Qtot_Vinf_2} 
\\
&
\mathcal{L}_{m}^{2}  \equiv   \,  
\frac{1}{4} \left\{\widehat{(UQ)}_m\right\}^2 
 \! + i (\Gamma_{L,m}-\Gamma_{R,m}) \, \widehat{(UQ)}_m^{}
-  \Delta_m^2   .
 \label{eq:L2_effective_def}
\end{align}
It is also deduced from 
Eqs.\ \eqref{eq:EOM_Qtot_Vinf}--\eqref{eq:EOM_Qtot_Vinf_2} that 
the second derivative of $p_{m}$ satisfies the equation 
\begin{align}
\frac{\partial^2 p_{m}^{}}{\partial t^2} \,= & \ 
- 4\, \mathcal{L}_{m}^{2} \,p_{m}^{} \;.
\label{eq:EOM_IL_Vinf}
\end{align}
The operator $\mathcal{L}_{m}^{2}$  plays a central role 
on the relaxation phenomena in the high-bias limit. 
Specifically, in the subspace where $\widehat{(UQ)}_m=0$,  
the eigenvalue of $\mathcal{L}_{m}^{2}$ is given simply by $-\Delta_m^2$, 
and the Heisenberg operators of  
$p_{m}$ and $q_{m}$ can be expressed 
as a linear combination of $e^{2 \Delta_m t}$ and $e^{-2 \Delta_m t}$. 
Here, the relaxation rate, $2 \Delta_m$, is determined by 
a damping of a particle-hole pair excitation.\cite{OguriSakano}
Furthermore, it can be deduced from these properties 
that in the high-bias limit a wide class of 
the susceptibilities of the charges and currents 
become identical to those for the noninteracting electrons 
 as shown in Sec.\ \ref{subsec:susceptibility}.

\section{Interaction representation 
for the non-Hermitian Hamiltonian}

\label{sec:boundary_condition_TFD}

We have introduced in the above  
$\widehat{H}_\mathrm{eff}^{}$ corresponds to  
the effective action $\mathcal{S}_0 + \mathcal{S}_U$. 
In order to complete the full description, 
we need to specify the density matrix that 
determines the nonequilibrium distribution.
Furthermore, it is also necessary 
to impose some conditions as  
the Fermion operators $d_{+,m}^{}$ of the $+$ branch  
describe the same physical particle 
as that of the $-$ branch  
at the turnaround point,  $t\to \infty$, of the Keldysh contour. 
%
As we see in the following, the time-dependent perturbation theory 
for the Liouville-Fock space can be constructed in a way such that 
these requirements can be fulfilled 
 through the properly chosen final $\langle \! \langle I |\! |$  
and initial $|\! | \rho \rangle \! \rangle$ states.\cite{TFD} 

To this end, 
we consider the time evolution in more detail 
in the interaction representation 
\begin{align}
\widehat{\mathcal{U}}(t_2,t_1) 
 \equiv& \  \mathrm{T} \exp  \left[ 
-i    \displaystyle 
\int_{t_1}^{t_2}     dt \, 
e^{i\widehat{H}_\mathrm{eff}^{(0)} t} 
\widehat{H}_\mathrm{eff}^{(U)}
e^{-i\widehat{H}_\mathrm{eff}^{(0)} t}
\right] , 
\label{eq:U_time_evolution} \\
\mathcal{O}^{\mathcal{I}}_{}(t) \,\equiv & \ \,  
  e^{i\widehat{H}_\mathrm{eff}^{(0)} t}\, 
  \mathcal{O}\,
  e^{-i\widehat{H}_\mathrm{eff}^{(0)} t} \;, 
\end{align}
where $\mathrm{T}$ is the usual time-ordering operator 
 along the branch of $-\infty < t < \infty$.

%

\subsection{Final and initial states:  
$\ \langle \! \langle I |\! |$ and  $|\! | \rho \rangle \! \rangle$ }
\label{subsec:interaction_representation}

The free part of the effective Hamiltonian 
can be rewritten in a diagonal form 
\begin{align} 
\widehat{H}_\mathrm{eff}^{(0)} 
\,=& \ 
\sum_m  \xi_{d,m}
\left( 
a_{m}^{-1} a_{m}^{} 
+
b_{m}^{-1} b_{m}^{} 
 -1  
\right) 
\nonumber \\
& \ +\,  
\sum_m \, i \Delta_m 
\left( 
a_{m}^{-1} a_{m}^{} 
-
b_{m}^{-1} b_{m}^{}
-1 \right)  \;.
\end{align} 
Here, 
$a_{m}^{-1}$ and $b_{m}^{-1}$
are defined with respect to the {\em left\/} eigenvectors 
of the non-Hermitian matrix $\bm{L}_{0,m}^{}$. 
Correspondingly,  $a_{m}^{}$ and $b_{m}^{}$ describe 
the {\em right\/} eigenvectors:
\begin{align} 
&
a_{m}^{} 
 \equiv    
\frac{ d_{-,m}^{} - d_{+,m}^{}
}{\sqrt{2}}
,  \ \ 
a_{m}^{-1} 
 \equiv 
\frac{\sqrt{2}\bigl(\Gamma_{L,m} d_{-,m}^{\dagger}
\! - \Gamma_{R,m} d_{+,m}^{\dagger}\bigr)}{\Gamma_{L,m}+\Gamma_{R,m}},   
\label{eq:Vinf_gamma_operators_1}
\\
&
b_{m}^{-1} 
 \equiv    
\frac{d_{-,m}^{\dagger} +  \, d_{+,m}^{\dagger}
}{\sqrt{2}} , 
\ \  
b_{m}^{} 
 \equiv 
\frac{\sqrt{2}\bigl(
\Gamma_{R,m} d_{-,m}^{\phantom{|}}
\! +  \Gamma_{L,m} d_{+,m}^{\phantom{|}} \bigr)}{\Gamma_{L,m}+\Gamma_{R,m}}.   
\label{eq:Vinf_gamma_operators_2}
\end{align} 
These operators satisfy the anti-commutation relations,    
\begin{align}
& 
\!\!
\bigl\{ a_{m}^{} , a_{m'}^{-1} \bigr\}  =  
\bigl\{ b_{m}^{} , b_{m'}^{-1} \bigr\} = \, \delta_{mm'} 
, \\
& 
\!\!
\bigl\{ b_{m}^{} , a_{m'}^{-1} \bigr\} = 
\bigl\{ a_{m}^{} , b_{m'}^{-1} \bigr\} = 
\bigl\{ a_{m}^{} , a_{m'\! } \bigr\} = 
\bigl\{ b_{m}^{} , b_{m'\! } \bigr\} = 0.
\end{align}
Since the eigenvalues of $\widehat{H}_\mathrm{eff}^{(0)}$ are  
complex, the corresponding eigenstates 
show a decaying or explosive long-time behavior for $t \to \infty$,  
\begin{align}
 a_{m}^{\mathcal{I}}(t) =  
 a_{m}^{} e^{(\Delta_m -i\xi_{d,m}) t},
\quad  \ 
 b_{m}^{\mathcal{I}}(t) =
 b_{m}^{} e^{-(\Delta_m+i\xi_{d,m}) t}  .  
\label{eq:a_b_interaction_rep}
\end{align}
The relaxation time is determined by  $\Delta_m$, i.e.,  
the imaginary part of the eigenvalue.
Thus, the final and initial states 
in the time-dependent perturbation theory  
for correlation functions 
 must satisfy a strong requirement that 
they should eliminate the explosive part, 
preserving only the decaying part. 
This condition is cleared by taking the states, in which  
all the explosive \lq\lq $a_{m}^{}$" particles are filled,   
as a set of \lq\lq vacuums"
\begin{align}
\langle \! \langle I |\! |  \,\equiv & \  
\left\langle 0 \right| 
a_{N}^{} \,a_{N-1}^{} \cdots\, a_{2}^{}\, a_{1}^{} 
\;, 
\label{eq:final_state}
\\
\qquad \quad 
|\! |  \rho  \rangle \! \rangle 
\,\equiv & \ \,  
a_{1}^{-1} \,a_{2}^{-1} \cdots \,a_{N-1}^{-1} \,a_{N}^{-1}
\left|0\right\rangle  \;.
\label{eq:initial_state}
\end{align}
These two states are normalized such that 
$\langle\!\langle I|\!| \rho \rangle \! \rangle = 1$.
We can see that the causal propagators defined with 
respect to these states correctly 
describe the relaxation process
\begin{align} 
\langle\!\langle I|\!|
\mathrm{T}\,
a^{\mathcal{I}}_{m}(t)\, a^{-1\mathcal{I}}_{m'}(0)
|\! | \rho \rangle \! \rangle 
\,=& \, - \delta_{mm'}\, \theta(-t)\,e^{(\Delta_m -i\xi_{d,m}) t} 
,
\label{eq:G0_aa_Vinf}
\\
\langle\!\langle I|\!|
\mathrm{T}\,
b^{\mathcal{I}}_{m}(t)\, b^{-1\mathcal{I}}_{m'}(0)
|\! | \rho \rangle \! \rangle 
\,= &  \ \ \  
\delta_{mm'} \,\theta(t)\, e^{-(\Delta_m+i\xi_{d,m}) t} 
\;.
\label{eq:G0_bb_Vinf}
\end{align}


%
The final state $\langle \! \langle I |\! |$ also satisfies  
the other requirement for the turnaround point of the Keldysh contour, 
\begin{align}
\langle \! \langle I |\! | d_{-,m}^{}
 =  \langle \! \langle I |\! | d_{+,m}^{}\;, \qquad
\langle \! \langle I |\! |  d_{-,m}^{\dagger} 
= - \langle \! \langle I |\! |   d_{+,m}^{\dagger} \;.
\label{eq:bra_boundary}
\end{align}
These relations hold for arbitrary $m$, and reproduce 
a linear dependence between the $-$ and $+$ components 
of the Keldysh correlation functions corresponding to  
Eqs.\ \eqref{eq:Gr_Keldysh_sum_rule} and 
\eqref{eq:Ga_Keldysh_sum_rule}.

In the final and initial states, 
defined in Eqs.\ 
\eqref{eq:final_state} and \eqref{eq:initial_state},
the charge $Q_{m}$ vanishes identically 
 for each $m$ of the orbitals,   
\begin{align}
& 
\langle\!\langle I|\!| Q_{m} \,=\, 0 , 
 \qquad \quad 
 Q_{m} |\!| \rho \rangle \! \rangle \,=\, 0 .   
\label{eq:Q=0} 
\end{align}
It can also be deduced from this property that 
 $\langle\!\langle I|\!|$ and $|\! | \rho \rangle \! \rangle$ are also 
the eigenstates for both,   
$\widehat{H}_\mathrm{eff}^{(0)}$  and  $\widehat{H}_\mathrm{eff}^{}$,
with  zero eigenvalue, 
\begin{align}
& 
\langle\!\langle I|\!| \widehat{H}_\mathrm{eff}^{(0)}\, = \,0\;,
  \qquad 
\langle\!\langle I|\!| \widehat{H}_\mathrm{eff}^{}\,=\,0\;, 
\label{eq:zero_mode_H_eff_bra} 
\\
& \  
\widehat{H}_\mathrm{eff}^{(0)} |\!|\rho \rangle \! \rangle \, = \,0\;,  
\qquad  
 \widehat{H}_\mathrm{eff}^{} |\!|\rho \rangle \! \rangle \,= \,0\;.   
\label{eq:zero_mode_H_eff_ket}
\end{align}
Therefore,  $\langle\!\langle I|\!|$ and $|\! | \rho \rangle \! \rangle$  
do not evolve in time in both the Schr\"{o}dginger 
and interaction representations, 
\begin{align}
&
\langle\!\langle I|\!|\, 
 e^{i\widehat{H}_\mathrm{eff}^{} t}
\,= \,  
\langle\!\langle I|\!|, 
\qquad 
\langle\!\langle I|\!|\,\widehat{\mathcal{U}}(t,t')  
\,= \,  
\langle\!\langle I|\!|, 
\\  
&
\!\!\!
 e^{-i\widehat{H}_\mathrm{eff}^{} t}
|\! | \rho \rangle \! \rangle   
\,=\, 
|\! | \rho \rangle \! \rangle ,
\qquad \ 
\widehat{\mathcal{U}}(t, t') 
|\! | \rho \rangle \! \rangle  
\, = \,   
|\! | \rho \rangle \! \rangle  . 
\label{eq:initial_final_no_evolve}
\end{align}
Specifically, in the interaction representation, 
the initial condition is given formally at $t\to -\infty$. 
Therefore, taking the initial condition to be     
 $|\! | \rho (-\infty) \rangle \! \rangle 
 \equiv  |\! | \rho \rangle \! \rangle$,   
the time evolution of the wavefunction 
in the interaction representation can be describe in the form,  
\begin{align}
&|\! | \rho (t) \rangle \! \rangle 
\, \equiv  \, 
\widehat{\mathcal{U}}(t, -\infty) 
|\! | \rho (-\infty)  \rangle \! \rangle 
\ =\, |\! | \rho \rangle \! \rangle \;.
\end{align} 
Then, the expectation values 
are defined with respect to the wavefunction at $t=0$   
that is the time the Heisenberg and the interaction 
representations coincide,
\begin{align}
\langle \mathcal{O}(t) \rangle  \equiv & \
\langle\!\langle I|\!| \mathcal{O} (t)|\!| \rho(0) \rangle \! \rangle 
\label{eq:average_Vinf}
\\
= & \ 
\langle\!\langle I|\!| 
\mathrm{T} \,\mathcal{O}^\mathcal{I} (t)
\, \widehat{\mathcal{U}}(\infty, -\infty) 
|\!| \rho(-\infty) \rangle \! \rangle 
\;.
\label{eq:average_Vinf_2}
\end{align}
This completes an explicit construction of the time-dependent 
perturbation theory for the high-bias limit.

\subsection{Statistical distributions at $eV \to \infty$}
\label{subsec:density_matrix_Vinf}

 The initial state $|\!| \rho \rangle\!\rangle$ 
determines the  density matrix 
in the limit of $eV \to \infty$, and 
has the properties similar to Eq.\ \eqref{eq:bra_boundary}, 
\begin{align}
 \Gamma_{L,m}\, d_{+,m}^{} |\!| \rho \rangle\!\rangle\, =&\ 
-\Gamma_{R,m}\, d_{-,m}^{} |\!| \rho \rangle\!\rangle\;,  
\label{eq:ket_boundary_1} \\
 \Gamma_{R,m} \,d_{+,m}^{\dagger} |\!| \rho \rangle\!\rangle \,= & \ \ 
\Gamma_{L,m}\, d_{-,m}^{\dagger} |\!| \rho \rangle\!\rangle \;. 
\label{eq:ket_boundary_2}
\end{align}
The underlying statistical weight can be extracted 
as a density matrix, defined such that  
$ \widehat{\rho}\,|\! |  I  \rangle \! \rangle  \equiv  
|\! |  \rho  \rangle$,\cite{TFD}  
\begin{align}
  \widehat{\rho} \, = & \,  
 \prod_{m=1}^N \bigl(\, 1 +  
2 r_{m} \, q_{m}
 \bigr) , 
\label{eq:density_matrix_Vinf} 
\quad  
r_{m} \equiv   \frac{\Gamma_{L,m}-\Gamma_{R,m}}{\Gamma_{L,m}+\Gamma_{R,m}} , 
\end{align}
 where  $|\! |  I  \rangle \! \rangle$ is 
a conjugate of $\langle \! \langle I |\! |$ 
whose explicit form is given in the right-hand side 
of Eq.\ \eqref{eq:final_state}.
This density matrix correctly describes 
the statistical distribution in the high-bias limit. 
Note that $\widehat{\rho}$ in this case does not depend 
on the interaction $U_{mm'}^{}$ 
but varies as a function of $r_m$ that parametrizes 
 the asymmetry in the dot-lead couplings.
Specifically, in the symmetric-coupling case where $r_m = 0$ for all $m$,   
it describes a uniform distribution 
 $\widehat{\rho} = 1$ and 
the average occupation of $n_{-,m}^{}$ 
becomes the same as that of $n_{+,m}^{}$.

The average formula  \eqref{eq:average_Vinf} 
reproduces exactly the local charges in the high-bias limit, 
\begin{align} 
\langle\!\langle I|\!| q_{m} |\!|\rho(0) \rangle \! \rangle 
\,= \, \frac{r_{m}}{2}\;,
\end{align}
and $\langle  n_{-,m}\rangle + \langle n_{+,m} \rangle = 1$.
Furthermore, the steady currents through the dot are 
also correctly reproduced,
\begin{align} 
 &
 \!\!\!\!
 \langle\!\langle I|\!| I_{R,m} |\!|\rho(0) \rangle \! \rangle 
  =  
 \langle\!\langle I|\!| I_{L,m} |\!|\rho(0) \rangle \! \rangle 
 \ = \   
 \frac{2\Gamma_{L,m}\Gamma_{R,m}}{\Gamma_{L,m}+\Gamma_{R,m}}\;.
 \end{align}
Note that the averages of the charges and currents 
do not depend on  $U_{mm'}^{}$ in the high-bias limit. 
 Similarly, 
the dynamic susceptibilities for charges and currents 
also take the noninteracting values 
as discussed in the next section.


\section{
Correlation functions in the 
thermal field theory for $eV \to\infty$
}

\label{sec:correlation functions}

In this section, we explain the relations between 
the Keldysh correlation functions 
and the corresponding thermal-field-theoretical ones. 
 We also describe some important high-bias properties.

\subsection{Dynamic susceptibility}
\label{subsec:susceptibility}

We consider a dynamic susceptibility, defined by  
\begin{align}
\chi^{\mu\nu}_{mm'}(t) 
\equiv  \, 
-i 
\sum_{\lambda\lambda'}
\bm{\tau}_3^{\mu\lambda}
\bm{\tau}_3^{\lambda'\nu}
\langle\!\langle I|\!| \,\mathrm{T}\, 
\delta n_{\lambda,m}(t) \,\delta  n_{\lambda',m'} 
|\!|\rho \rangle \! \rangle  ,
\label{eq:correlation_function_physical_charge}
\end{align}
where $\delta n_{\mu,m} \,\equiv\, n_{\mu,m} 
- \langle n_{\mu,m} \rangle$ for $\mu =-,+$.
The Pauli matrix $\bm{\tau}_3$ has been multiplied 
so that 
each  ($\mu$, $\nu$) component of $\chi^{\mu\nu}_{mm'}$
coincides with the corresponding element of the Keldysh susceptibility.

Equation \eqref{eq:correlation_function_physical_charge} 
 can be calculated further,  
rewriting it in terms of the relative charge   
$\delta q_{m}^{} \equiv q_{m}^{} - \langle q_{m}^{} \rangle$,  
\begin{align}
\chi^{\mu\nu}_{mm'}(t) 
 =& \
-i \langle\!\langle I|\!| \,\mathrm{T}\,  
\delta q_{m}^{}(t) \,
\delta q_{m'}^{} (0)
|\!|\rho \rangle \! \rangle 
\nonumber \\
 =& \ 
-i \theta(t) \, 
\langle\!\langle I|\!| \,\delta q_{m}^{}\,
e^{-i\widehat{H}_\mathrm{eff}^{} t}
\delta q_{m'}^{}|\!|\rho \rangle \! \rangle 
\nonumber \\
& \ -i \theta(-t) \, 
\langle\!\langle I|\!| \,\delta q_{m'}^{}\,
e^{i\widehat{H}_\mathrm{eff}^{} t}
\delta q_{m}^{}|\!|\rho \rangle \! \rangle 
\;.
\label{eq:correlation_function_physical_charge2}
\end{align}
We have used an identity $\delta n_{\mu,m}  = 
 \frac{1}{2}\,Q_{m} - \mathrm{sign}(\mu) \,\delta q_{m}^{}$   
and the high-bias properties described in 
Eq.\ \eqref{eq:Q=0} to obtain the second line. 
This expression shows that the dynamics of the excited states   
$\langle\!\langle I|\!| \,\delta q_{m}^{}$ 
and 
$\delta q_{m'}^{}|\!|\rho \rangle \! \rangle$ 
determine the time evolution of $\chi^{\mu\nu}_{mm'}(t) $.
In these two states one particle-hole pair is excited, respectively, 
from the \lq\lq vacuums"  
 $\langle\!\langle I|\!|$ and  $|\!|\rho \rangle \! \rangle$,     
by the matrix elements  
\begin{align}
\delta q_{m}^{} 
= &  \  
\frac{1}{2}  
\left( 
 a_{m}^{-1}b_{m}^{} 
+ \frac{4\Gamma_{L,m} \Gamma_{R,m}}{\Delta_m^2}\, b_{m}^{-1} a_{m}^{} 
\right)
\nonumber \\
& + 
\frac{r_m}{2}\left( a_{m}^{-1} a_{m}^{} - b_{m}^{-1} b_{m}^{} -1 \right) \;, 
\end{align}
as 
\begin{align}
&\langle\!\langle I|\!| \,\delta q_{m}^{}
 \, = \  \frac{1}{2}\,\left\langle 0 \right| 
a_{N}^{} \, \cdots\,a_{m+1}^{} \,b_{m}^{}
\,a_{m-1}^{}\cdots\, a_{1}^{} 
,
\label{eq:intermediate_state_susceptibilities_1}
\\
&\delta q_{m}^{}|\!|\rho \rangle \! \rangle 
\,  = \ 
\frac{2\Gamma_{L,m} \Gamma_{R,m}}{\Delta_m^2}\   
a_{1}^{-1} \cdots \,a_{m-1}^{-1} \,b_{m}^{-1}
\,a_{m+1}^{-1} \cdots\, \,a_{N}^{-1}
\left|0\right\rangle  .
\label{eq:intermediate_state_susceptibilities_2}
\end{align}
The particle-hole pair excitation does not change 
the total $Q_{m}$ in the Liouville-Fock space, and thus 
\begin{align}
& \langle\!\langle I|\!| \delta q_{m}^{}\,Q_{m''}\,=\,0 \;, \qquad 
 Q_{m''}\,\delta q_{m'}^{}|\!|\rho \rangle \! \rangle 
\,=\,0 \;,
\label{eq:sus_Q=0}
\end{align}
for all $m''$ ($=1,2,\ldots,N$). 
Therefore, in 
Eq.\ \eqref{eq:correlation_function_physical_charge2} 
the operators $Q_{m''}$'s  
included in $\widehat{H}_\mathrm{eff}^{}$ 
 in the intermediate states can be  
replaced by the corresponding eigenvalues, $Q_{m''}=0$ for all $m''$,  
\begin{align}
& 
\!\!\!\!
\chi^{\mu\nu}_{mm'}(t) \, =  
\nonumber \\
& \ 
-i \theta(t) \, 
\langle\!\langle I|\!| \,\delta q_{m}^{}\,
e^{-i\sum_{m''}\left( 
\bm{d}_{m''}^{\dagger} \bm{L}_{0,m''}^{} \bm{d}_{m''}^{}  - i\Delta_{m''} 
\right)  t}
\delta q_{m'}^{}|\!|\rho \rangle \! \rangle 
\nonumber \\
& \ 
 -i \theta(-t) \, 
\langle\!\langle I|\!| \,\delta q_{m'}^{}\,
e^{i \sum_{m''}\left( 
\bm{d}_{m''}^{\dagger} \bm{L}_{0,m''}^{} \bm{d}_{m''}^{}  - i\Delta_{m''} 
\right)  t}
\delta q_{m}^{}|\!|\rho \rangle \! \rangle  ,
\label{eq:correlation_function_physical_charge3}
\end{align}
and thus the interaction term vanishes in the intermediate states, 
Consequently, 
 the dynamic susceptibility is asymptotically free in the high-bias limit 
as it coincides with the noninteracting form,
\begin{align}
\chi^{\mu\nu}_{mm'}(t)  = &  \,  
-i\, \delta_{mm'}  \, \frac{\Gamma_{L,m} \Gamma_{R,m}}{(\Gamma_{L,m}+\Gamma_{R,m})^2} 
\ e^{-2\Delta_m |t|} 
\;.
 \label{eq:qq_corelation}
\end{align}
Alternatively, one can calculate 
 $\chi^{\mu\nu}_{mm'}(t)$ from the equation of motion, 
in which the decay rate $2\Delta_m$ appears 
as the eigenvalue of $\mathcal{L}_{m}^{2}$, 
mentioned in Sec.\ \ref{subseec:current_algebra}. 

This asymptotically-free behavior in the high-bias limit is common to 
  a wide class of the correlation functions $X_{AB}^{}(t,t')$, 
defined with respect to the operators $A$ and $B$ 
 which commute with $Q_m$ for all $m$; 
\begin{align}
X_{AB}^{}(t,t')  
\,\equiv&  \ 
-i\, 
\langle\!\langle I|\!| \,\mathrm{T}\, 
A(t) \,
B(t') 
|\!|\rho(0) \rangle \! \rangle \; ,
\label{eq:correlation_function}
\\
\bigl[A\,,\, Q_m \bigr]\, = & \  
\bigl[B\,,\, Q_m \bigr] \, =\,0 \;.
\label{eq:Q_conservation_A_B}
\end{align}
For this correlation function, the relations corresponding  
to Eq.\ \eqref{eq:sus_Q=0} follow for both $A$ and $B$  
from the condition  \eqref{eq:Q_conservation_A_B},
and thus the interaction effects vanish as that 
in the case of the dynamic susceptibility $\chi^{\mu\nu}_{mm'}(t)$. 
One important example of this is the shot noise 
that can be derived from the current-current correlation function.   
Because the current operator $I_{\alpha,m}^{}$ satisfies 
a commutation relation   
$\bigl[I_{\alpha,m}^{} \,,\, Q_m \bigr]  =0$, 
the high-bias asymptotic form of  
the $\omega$-dependent current fluctuations becomes identical to 
the noninteracting result also in the multi-orbital case 
as that in the $N=2$ case.\cite{OguriSakano}


\subsection{Green's function}
\label{subsec:TFD_green}

We next describe the correspondence between the Keldysh Green's function 
and the ones defined with respect to the Liouville-Fock space.   
The free Green's function for $\widehat{H}_\mathrm{eff}^{(0)}$ 
is defined by 
\begin{align} 
 \mathcal{G}_{0,m}^{\mu\nu}(t) 
\, \equiv  & \  - i \,
\langle\!\langle I|\!|
\mathrm{T}\,
d^{\mathcal{I}}_{\mu,m}(t)\, d^{\dagger\mathcal{I}}_{\nu,m}(0)
|\! | \rho \rangle \! \rangle \;.
\label{eq:G_eff_Vinf_0}
\end{align}
This function can be calculated,    
using Eqs.\ \eqref{eq:a_b_interaction_rep}--\eqref{eq:final_state},  
\begin{align}
& \bm{\mathcal{G}}_{0,m}^{}  
  \equiv   
 \left[\, 
  \begin{matrix}
   \mathcal{G}^{--}_{0,m} & \mathcal{G}^{-+}_{0,m}   \cr
   \mathcal{G}^{+-}_{0,m} & \mathcal{G}^{++}_{0,m}  \cr  
  \end{matrix} 
  \, \right]  
\  = \   
\bm{G}_{0,m}^{} \, \bm{\tau}_3 \;.
 \label{eq:G0_TFD_Vinf} 
\end{align}
Here, $\bm{G}_{0,m}(\omega)$ is the high-bias asymptotic form of the 
Keldysh Green's function
 \begin{align}
 & \left\{\bm{G}_{0,m}(\omega)\right\}^{-1} 
 = \, \bm{\tau}_3^{} \,\Bigl[\,  
(\omega-\xi_{d,m}) \bm{1} 
 \, - \,\bm{L}_0 \, \Bigr]\;.
 \label{eq:G0_keldysh_rev} 
\end{align}
Note that  $\varepsilon_{d,m}$ which appeared 
in the original definition of $\bm{G}_{0,m}(\omega)$ 
 given in Eq.\ \eqref{eq:G0_keldysh} 
has been replaced by $\xi_{d,m}$, including 
the energy shift defined in Eq.\ \eqref{eq:Xi} into 
the non-perturbed part. 


The interacting Green's function for the Liouville-Fock space is 
defined by
\begin{align} 
 \mathcal{G}_{m}^{\mu\nu}(t) 
 \equiv  & \  - i \,
\langle\!\langle I|\!|
\mathrm{T}\,
d^{\phantom{\dagger}}_{\mu,m}(t)\, d^{\dagger}_{\nu,m}(0)
|\! | \rho(0) \rangle \! \rangle 
\label{eq:G_eff_Vinf}
\\
 =  & \,  - i \,
\langle\!\langle I|\!|
\mathrm{T}\,
d^{\mathcal{I}}_{\mu,m}(t)\, d^{\dagger\mathcal{I}}_{\nu,m}(0)
\,\widehat{\mathcal{U}}(\infty, -\infty) 
|\! | \rho \rangle \! \rangle \;.
\label{eq:G_eff_Vinf_interaction_representation}
\end{align}
The same relation holds between the interacting 
Green's functions,
 $\bm{\mathcal{G}}_{m}^{}$ and $\bm{G}_{m}^{}$, 
as that in the noninteracting case 
\begin{align}
\bm{\mathcal{G}}_{m}^{}  
\,=& \  
\bm{G}_{m}^{} 
\, \bm{\tau}_3 
\;. 
\label{eq:GKldysh_Gtfd}
\end{align}
This can be verified perturbatively, using    
the Feynman diagrammatic expansion      
which can be generated from $\widehat{\mathcal{U}}(\infty, -\infty)$  
defined in Eq.\ \eqref{eq:G_eff_Vinf_interaction_representation}. 
The noninteracting Green's function  $\bm{\mathcal{G}}_{0,m}^{}$
that is assigned to the Feynman diagrams 
has  one-to-one correspondence with the Keldysh propagator 
$\bm{G}_{0,m}^{}$, as shown in Eq.\ \eqref{eq:G0_TFD_Vinf}.  
Furthermore, the Feynman rule 
for $\bm{\mathcal{G}}_{m}^{}$ is  
essentially the same as that for $\bm{G}_{m}^{}$ 
in the Keldysh formalism. 
There is a slight difference in the treatment of 
the Hartree term but the counter term, 
which is a part of $\widehat{H}_\mathrm{eff}^{}$, 
compensates the difference 
as shown in Appendix \ref{sec:Hartree}.
Therefore, there is an exact diagram to diagram 
correspondence between the Keldysh and thermal-field-theoretical  
perturbation expansion, and thus Eq.\ \eqref{eq:GKldysh_Gtfd} holds. 
Note that the sign that arises from $\bm{\tau}_3$ also appears in the 
relation between  
the Keldysh self-energy $\bm{\Sigma}_{m}^{}$
for $\bm{G}_{m}$ and the corresponding self-energy 
$\bm{\mathit{\Sigma}}_{m}^\mathrm{TFT}$, 
defined by $\{ \bm{\mathcal{G}}_{m}\}^{-1}= 
\{ \bm{\mathcal{G}}_{0,m}\}^{-1}-\bm{\mathit{\Sigma}}_{m}^\mathrm{TFT}$, 
as   
\begin{align}
\bm{\Sigma}_{m}^{}(\omega)
\, = \, \bm{\tau}_3 \,\bm{\mathit{\Sigma}}_{m}^\mathrm{TFT}(\omega)\;.
\end{align}

The four components of $\mathcal{G}_{m}^{\mu\nu}$
have the same linear dependence as that  
the Keldysh components $G_{m}^{\mu\nu}$ have.  
Therefore, the retarded $G_{m}^r$ and advanced $G_{m}^a$ 
Green's functions can be expressed in two different forms, 
\begin{align}
G_{m}^r 
\,=& \  
\mathcal{G}_{m}^{--} + \mathcal{G}_{m}^{-+}
\,=\, 
\mathcal{G}_{m}^{+-} + \mathcal{G}_{m}^{++} \;, 
\label{eq:Gr_Keldysh_sum_rule}
\\
G_{m}^a 
\,=& \  
\mathcal{G}_{m}^{--} - \mathcal{G}_{m}^{+-}
\,=\, 
\mathcal{G}_{m}^{++} - \mathcal{G}_{m}^{-+} \;.
\label{eq:Ga_Keldysh_sum_rule}
\end{align}
In the high-bias limit,  
 $\bm{\mathcal{G}}_{m}$ can be expressed in terms of 
 these two Green's functions  
\begin{align}
\!\!
\bm{\mathcal{G}}_{m}^{}
\,=\,    
 \frac{G_{m}^r}{\Delta_m}
 \left [ 
\begin{matrix} 
\Gamma_{R,m} & \Gamma_{L,m}\cr 
\Gamma_{R,m} & \Gamma_{L,m}
 \rule{0cm}{0.5cm}  \cr  
\end{matrix}          
 \right ] 
 +  
 \frac{G_{m}^a}{\Delta_m}
 \left [  
\begin{matrix} 
\ \Gamma_{L,m} & -\Gamma_{L,m}\cr 
-\Gamma_{R,m} & \ \Gamma_{R,m}
 \rule{0cm}{0.5cm}  \cr  
\end{matrix}          
 \right ] .
\label{eq:G_Vinf_intermediate_form_in_time}
\end{align} 
This is because the statistical distribution 
 for $eV \to \infty$ is determined by a time-independent 
state $|\! | \rho \rangle \! \rangle$ 
as shown in Eqs.\ \eqref{eq:ket_boundary_1}--\eqref{eq:density_matrix_Vinf}.  
Furthermore, 
only a single component among the four is independent 
since the relation $G_{m}^a(\omega)= 
\left\{G_{m}^r(\omega)\right\}^*$ 
holds in the frequency representation.  
For this reason,  
we consider mainly the retarded Green's function 
 in the rest of the paper.

\begin{widetext}

\section{Exact interacting Green's function for $eV \to \infty$}
\label{sec:calculations_for_G}

In this section, 
we describe a derivation of the asymptotic 
form of Green's function in the high-bias limit.

\subsection{Generic form in the high-bias limit}

The retarded Green's function can be  expressed 
in the following form,
using Eq.\ \eqref{eq:Gr_Keldysh_sum_rule},
\begin{align}
G_{m}^r (t) 
\,= & \ 
\frac{1}{2} \Bigl(\, \mathcal{G}_{m}^{--}(t) + \mathcal{G}_{m}^{-+}(t)
+ \mathcal{G}_{m}^{+-}(t) + \mathcal{G}_{m}^{++}(t) \,\Bigr) \;, 
\\
\,=& \   
-i \,\theta(t)\, \frac{1}{2}\,
\langle\!\langle I|\!| 
\left(
d_{-, m}^{\phantom{\dagger}}(t) 
+ d_{+, m}^{\phantom{\dagger}}(t) 
\right)
\, 
\left(
d_{-,m}^{\dagger} + 
d_{+,m}^{\dagger}
\right)
|\! |  \rho(0)  \rangle \! \rangle   
\;.
\label{eq:retarded_G_V_inf}
\end{align}
This can be rewritten further,  
using the properties of 
 $\langle\!\langle I|\!|$ and $|\! | \rho \rangle \! \rangle$ 
given in Eqs.\ 
\eqref{eq:zero_mode_H_eff_bra}--\eqref{eq:zero_mode_H_eff_ket},  
\begin{align}
G_{m}^r (t) 
\,=  \,   
-i \,\theta(t) \  
\langle\!\langle I_m^{}|\!| \,
e^{-i\widehat{H}_\mathrm{eff}^{} t}
\,|\! |  \rho_m^{}  \rangle \! \rangle   
\;.
\label{eq:retarded_G_V_inf2}
 \end{align}
Here,  
 $\langle\!\langle I_m^{}|\!|$ and $|\! |  \rho_m^{}  \rangle \! \rangle$ 
denote the intermediate states  with 
single-particle excitations,   
\begin{align}
\langle \! \langle I_m^{} |\! |  \,\equiv & \ \, 
\langle \! \langle I |\! |\, 
\frac{1}{\sqrt{2}}
\left(
d_{-, m}^{\phantom{\dagger}}+ d_{+, m}^{\phantom{\dagger}} 
\right) 
\ = \ 
(-1)^{m-1}\, 
\left\langle 0 \right| 
a_{N}^{} \, \cdots\,a_{m+1}^{} \,d_{-,m}^{}d_{+,m}^{}
\,a_{m-1}^{}\cdots\, a_{1}^{} 
\;,  \\
|\! |  \rho_m^{}  \rangle \! \rangle 
\,\equiv & \ \,  
\frac{1}{\sqrt{2}}
\left(
d_{-,m}^{\dagger} + 
d_{+,m}^{\dagger}
\right)
|\! |  \rho  \rangle \! \rangle 
\ = \ 
  (-1)^{m-1} \ \,  
a_{1}^{-1} \cdots \,a_{m-1}^{-1} \,d_{+,m}^{\dagger}d_{-,m}^{\dagger}  
\,a_{m+1}^{-1} \cdots\, \,a_{N}^{-1}
\left|0\right\rangle 
\;.
\label{eq:initial_final_states_dash}
\end{align}
In contrast to the particle-hole pair excitation 
for the dynamic susceptibilities described in Eqs.\ 
\eqref{eq:intermediate_state_susceptibilities_1}-\eqref{eq:intermediate_state_susceptibilities_2}, 
in the single-particle states  
$\langle\!\langle I_m^{}|\!|$ and $|\! |  \rho_m^{}  \rangle \! \rangle$  
the orbital $m$ is doubly occupied 
while  all the 
other orbitals $m'$ ($\neq m$) are kept unchanged in a similar way. 
Thus, for $\widehat{H}_\mathrm{eff}^{}$ which determines 
time evolution of the intermediate state 
described in Eq.\ \eqref{eq:retarded_G_V_inf2}, 
the operators $Q_{m'}$'s  
 can be replaced by their eigenvalues; 
$Q_{m} = 1$ and  $Q_{m'}=0$ for $m' \neq m$.  
This significantly simplifies Eq.\ \eqref{eq:retarded_G_V_inf2}, 
and makes the correlation effects factorizable in a bilinear form 
\begin{align}
G_{m}^r (t) 
= & \  
-i \,\theta(t) \  
\, 
e^{-i(\xi_{d,m}-i\Delta_m)  t}  
\prod_{m'(\neq m)}
\! e^{-\Delta_{m'}  t}  \, 
\langle\!\langle I_m^{}|\!| \,
e^{-i  \,\bm{d}_{m'}^{\dagger} \widetilde{\bm{L}}_{m'}^{(m)} \bm{d}_{m'}^{} t}
\,|\! |  \rho_m^{}  \rangle \! \rangle   
\;, 
\label{eq:retarded_G_V_inf3}
\\
\widetilde{\bm{L}}_{m'}^{(m)} 
 \equiv &   \ 
i\left [ \,
\begin{matrix} 
 \Gamma_{L,m'} - \Gamma_{R,m'} 
-i\,\frac{\displaystyle\mathstrut 1}{\displaystyle\mathstrut 2}\,U_{m'm}^{}  
& -2\Gamma_{L,m'}\,  \cr
 -2\Gamma_{R,m'}  
&  -(\Gamma_{L,m'} - \Gamma_{R,m'}) 
+i\,\frac{\displaystyle\mathstrut 1}{\displaystyle\mathstrut 2}\,U_{m'm}^{}      
\rule{0cm}{0.5cm}  
\cr
\end{matrix}          
\right] .
\end{align}
%
The matrix $\widetilde{\bm{L}}_{m'}^{(m)}$ consists of the free part  
$\bm{L}_{0}^{}$ defined in Eq.\ \eqref{eq:L0_matrix_Vinf} 
and the correction due to the inter-electron interaction.
The product in Eq.\ \eqref{eq:retarded_G_V_inf3} can be calculated 
separately for each $m'$ ($\neq m$), as 
\begin{align}
\langle\!\langle I_m^{}|\!| \,
e^{-i  \,\bm{d}_{m'}^{\dagger} \widetilde{\bm{L}}_{m'}^{(m)} \bm{d}_{m'}^{} t}
\,|\! |  \rho_m^{}  \rangle \! \rangle   
\, = & \  
 \left [ \, 
\begin{matrix} 
1   &  -1 \cr
\end{matrix}          
\, \right ] 
e^{-i  \, \widetilde{\bm{L}}_{m'}^{(m)} t}
 \left [  
\begin{matrix} 
 \ \frac{\Gamma_{L,m}}{\Delta_m}   \cr  
-  \frac{\Gamma_{R,m}}{\Delta_m}   \rule{0cm}{0.4cm} \cr
\end{matrix}          
\, \right ]  
\ \ = \ 
Z_{m'}^{(m+)}\, e^{-i\mathcal{E}_{m'}^{(m)}t}    
+Z_{m'}^{(m-)}\,  e^{i\mathcal{E}_{m'}^{(m)}t} \;.
\label{eq:Z_expression}
\end{align}
Here, $\mathcal{E}_{m'}^{(m)}$ 
is a complex eigenvalue of $\widetilde{\bm{L}}_{m'}^{(m)}$ 
and $Z_{m'}^{(m\pm)}$ is a weight factor 
determined by the corresponding eigenvector,
\cite{SaptosovWegewijs}
\begin{align}
 Z_{m'}^{(m\pm)} \,  \equiv & \       
\frac{1}{2} \left( 1 \pm 
\frac{i\Delta_{m'} + \frac{r_{m'}}{2}\,U_{m'm}^{} }{\mathcal{E}_{m'}^{(m)}}
\right), \qquad \qquad 
\mathcal{E}_{m'}^{(m)}\, 
\, \equiv \,  \sqrt{\frac{1}{4}\, U_{m'm}^{2} \!\! 
\, -  \Delta_{m'}^2  + i r_{m'} \Delta_{m'}\,  U_{m'm}^{} } \; . 
\label{eq:Z_and_E_for_general}
\end{align}
We obtain the explicit expression of the retarded Green's function, 
substituting Eq.\ \eqref{eq:Z_expression}
into Eq.\ \eqref{eq:retarded_G_V_inf3}, 
\begin{align}
%
 G_{m}^r(t) 
\,=& \     
-i\,\theta(t) 
\, e^{-i(\xi_{d,m}-i\Delta_m)  t}  
\prod_{m'(\neq m)}
\! e^{-\Delta_{m'}  t}  \, 
\Bigl(
 Z_{m'}^{(m+)}\, e^{-i\mathcal{E}_{m'}^{(m)}t}    
+Z_{m'}^{(m-)}\,  e^{i\mathcal{E}_{m'}^{(m)}t} 
\Bigr) \;.
\label{eq:retarded_G_V_inf4}
\end{align}
\end{widetext}
This is a main result of the present work, 
and the  Green's function can be written in  
a factorized form in the time representation. 

The asymptotically exact result for $eV \to \infty$ 
 captures essential physics of relaxation  
of interacting electrons at high energy scales.
The imaginary part of $\mathcal{E}_{m'}^{(m)}$ is bounded 
in the range $|\mathrm{Im}\, \mathcal{E}_{m'}^{(m)}| \leq \Delta_{m'}$,
and it certifies that $G_{m}^r(t)$ decays at long time.\cite{SaptosovWegewijs} 
The squared eigenvalue,  $\{\mathcal{E}_{m'}^{(m)}\}^2$,  also corresponds to  
the eigenvalue of the operator $\mathcal{L}_{m}^{2}$ 
for  $\widehat{(UQ)}_m = U_{m'm}$, which is defined 
in Eq.\  \eqref{eq:L2_effective_def}. 
This means that 
the particle-hole-pair excitation in  
the intermediate state evolves in time and 
contributes to the relaxations, which we can see more clearly 
in the continued-fraction representation in the next section.
Note that 
the high-bias expression Eq.\ \eqref{eq:retarded_G_V_inf4} 
in the symmetric-coupling case, where $r_{m''}=0$ for all $m''$,     
can also be regarded as 
an exact high-temperature Green's function at equilibrium 
because of the relation described in Eq.\ \eqref{eq:f_eff_high_energy}.      
The Fourier transform 
$G_{m}^r(\omega) =\int_0^{\infty} \! dt \, 
e^{i (\omega+i0^+) t}\, G_{m}^r(t)$,   
which can be carried out by expanding the product, 
becomes  a function of  $\omega_m \equiv \omega -\xi_{d,m}$ 
in the frequency representation. Alternatively, 
it can also be calculated, using a resolvent form of 
Eq.\ \eqref{eq:retarded_G_V_inf2}, 
\begin{align}
G_{m}^r (\omega) 
\,=  \,   
\langle\!\langle I_m^{}|\!| \ 
\frac{1}{\omega - \widehat{H}_\mathrm{eff}^{} +i0^+}
\ |\! |  \rho_m^{}  \rangle \! \rangle   
\;.
\label{eq:retarded_G_V_inf5}
\end{align}

\subsection{Some special cases}
\label{subsec:special_limits}

We examine some special cases in this subsection. 
The first one is the free-particle limit 
where $\widehat{H}_\mathrm{eff}^{(U)} \to 0$.   
Equation \eqref{eq:retarded_G_V_inf4} obviously  
reproduces the free propagator   
\begin{align}
 G_{0,m}^r(t) \,=\,
-i\,\theta(t) 
\, e^{-i(\xi_{d,m}-i\Delta_m)  t}  \;, 
\end{align}
as $e^{-\Delta_{m'}  t} \langle\!\langle I_m^{}|\!| \,
e^{-i  \,\bm{d}_{m'}^{\dagger} \widetilde{\bm{L}}_{m'}^{(m)} \bm{d}_{m'}^{} t}
\,|\! |  \rho_m^{}  \rangle \! \rangle \to 1 $ for $m'\neq m$ 
in the noninteracting case.

The second example is the case where one of the two leads are disconnected. 
In the limit $r_{m''}\to +1$ ($-1$) for all $m''$, 
the right (left) lead is disconnected,  
and the impurity level with 
the width $\Delta_m \to \Gamma_{L,m}$  ($\Gamma_{R,m}$) 
is fully occupied (empty). 
Then,  the Green's function takes the form 
\begin{align}
\lim_{\{r_{m''}\} \to \pm 1 } 
\! G_{m}^r(t) 
=     
-i\,\theta(t) 
\, e^{-i\left( \xi_{d,m} 
\pm \frac{N-1}{2} \overline{U}_m -i\Delta_m 
 \right)\,  t}  . 
\label{eq:limit_r_pm1}
\end{align}
The corresponding spectral function  for $r_{m''}\to \pm 1$ 
has a single Lorentzian peak 
at $\omega = \xi_{d,m} \pm (N-1)\overline{U}_m/2$
with $\overline{U}_m \equiv \sum_{m'(\neq m)} U_{m'm}/(N-1)$.  
Note that the peak position depends on which of the leads,  
$R$ or $L$, is disconnected. 

 The third one is  the atomic limit, where both  $\Gamma_{L,m''}$ and 
$\Gamma_{R,m''}$ vanish for all $m''$. 
In this case, the complex eigenvalue and weight factors approach   
$\mathcal{E}_{m'}^{(m)}\to U_{m'm}/2 $ and $Z_{m'}^{(m\pm)}\to 1/2$.  
Then,  Eq.\ \eqref{eq:retarded_G_V_inf4} takes the form 
\begin{align}
& \lim_{\{\Gamma_{R,m''}\} \to 0 \atop
\{\Gamma_{L,m''}\} \to 0 } G_{m}^r(t) 
\nonumber \\
& = \,     
-i\,\theta(t) 
\, e^{-i\xi_{d,m}  t}  
\prod_{m'(\neq m)}
\frac{
 e^{-i\frac{1}{2}U_{m'm}t}    
+ e^{i\frac{1}{2}U_{m'm}t}
}{2} \;.
\end{align}

\begin{widetext}

\section{
Green's function for the uniform interaction case}
\label{sec:uniform_interaction}

In this section, we consider the high-bias Green's function  
for the  $m$ independent interactions and hybridizations, 
choosing  $U_{mm'}^{} = U$, $\Delta_m = \Delta$, and $r_m = r$ 
for all $m$ and $m'$. 
However, the impurity levels $\xi_{d,m} = \varepsilon_{d,m}+(N-1)U/2$ 
can still be dependent on $m$,  
and also the coupling can be asymmetric $\Gamma_L \neq \Gamma_R$. 
Then, Eq.\ \eqref{eq:retarded_G_V_inf4} takes the form, 
\begin{align}
& G_{m}^r(t) 
=    
-i\,\theta(t) 
\, e^{-i(\xi_{d,m}-iN\Delta ) t}   
\left(\,
 Z_{}^{(+)}\, e^{-i\mathcal{E}t}    
+Z_{}^{(-)}\,  e^{i\mathcal{E}t} 
\,\right)^{N-1} .
\label{eq:retarded_G_V_inf_SUn_time}
\end{align}
Here, $\mathcal{E}$ and $Z_{}^{(\pm)}$ 
correspond to $\mathcal{E}_{m'}^{(m)}$ and 
$ Z_{m'}^{(m\pm)}$ 
for the uniform parameters, respectively.     
This Green's function can be rewritten in a partial fraction 
form, carrying out the Fourier transform using the binominal expansion,
\begin{align}
G_{m}^r(\omega) 
 \,=  \  \sum_{\mathcal{Q}=0}^{N-1} 
 \left( 
\begin{matrix} 
 N-1   \cr  
 \mathcal{Q} \cr
\end{matrix}          
\right) 
\frac{\left\{Z^{(+)}\right\}^\mathcal{Q} \left\{Z^{(-)}\right\}^{N-1-\mathcal{Q}}}{\omega-\xi_{d,m}+iN\Delta+(N-1-2\mathcal{Q})\,\mathcal{E}} 
\;.
\label{eq:retarded_G_V_inf_SUn_partial_fraction}
\end{align}
Note that 
the imaginary part of $G_{m}^r(\omega)$ 
is determined not only by $i N\Delta$ in the denominator   
but also through the complex parameters $\mathcal{E}$ and $Z^{(\pm)}$.

\subsection{Continued fraction representation} 

The Green's function can also be expressed 
in a continued fraction form, converting   
Eq.\ \eqref{eq:retarded_G_V_inf_SUn_partial_fraction} 
or carrying out the Householder transformation 
for Eq.\ \eqref{eq:retarded_G_V_inf5}, 
\begin{align}
G_{m}^r(\omega) =\, 
\cfrac{1}{\omega_m - \mathcal{A}_1 \frac{rU}{2} +i\, \mathcal{C}_1 \Delta  - 
\cfrac{\mathcal{B}_1\,(1-r^2)\left(\frac{U}{2}\right)^2}
{\omega_m - \mathcal{A}_2 \frac{rU}{2} + i\, \mathcal{C}_2 \Delta -    
\cfrac{\mathcal{B}_2\,(1-r^2)\left(\frac{U}{2}\right)^2}
{ \ \  \ddots \ - 
\cfrac{\ddots}  
{\omega_m -\mathcal{A}_{N-1} \frac{rU}{2}+ i\, \mathcal{C}_{N-1} \Delta -  
\cfrac{\mathcal{B}_{N-1}\,(1-r^2)\left(\frac{U}{2}\right)^2}
{\omega_m - \mathcal{A}_N \frac{rU}{2} + i\, \mathcal{C}_N  \Delta
}}}}} 
\;,
\label{eq:retarded_G_V_inf_SUn_continued_fraction}
\end{align}
where $\omega_m =\omega - \xi_{d,m}$.
The square-root dependence due to $\mathcal{E}$ disappears  
in the continued fraction representation as  
the coefficients $\mathcal{A}_k$, $\mathcal{B}_k$, and $\mathcal{C}_k$  
(for $k=1,\,2,\,\ldots\,, N$) 
are integers, which  do not depend on the physical parameters  
\begin{align}
\mathcal{A}_k =\, N-1 -2(k-1)  
\;, \qquad
\mathcal{B}_k =\, k\, (N-k) \;, \qquad 
\mathcal{C}_k =\, 2k-1  \;.
\label{eq:A_B_C_coefficients}
\end{align}
 The coefficient $\mathcal{A}_k$, which determines 
the energy shifts due to the coupling asymmetry, 
decreases as $k$ increases and changes 
the sign at the middle of $k$ between $1$ and $N$. 
The coefficient $\mathcal{B}_k$  
corresponds to the residue of intermediate states with $k$ particle-hole 
pairs, and  has a maximum at the middle of $k$. 
In contrast, the coefficient $\mathcal{C}_k$ increases linearly with $k$.  
It determines the relaxation rate in the high-bias limit,   
and can be  decomposed into two parts 
 $\mathcal{C}_k \Delta  = (k-1) 2\Delta + \Delta$.
The first term can be interpreted as a sum of 
the damping rate of $k-1$ intermediate particle-hole pairs 
each of which decays 
with the ratio of $2\Delta$ as mentioned in Sec.\ 
\ref{subseec:current_algebra} and \ref{subsec:susceptibility}, 
and the second term $\Delta$ corresponds 
to the decay rate of the single incident particle. 
Note that the initial part of the continued fraction, 
Eq.\ \eqref{eq:retarded_G_V_inf_SUn_continued_fraction},   
can be expressed in the  form 
\begin{align}
&
\omega_m - \mathcal{A}_1 \frac{rU}{2} +i \,\mathcal{C}_1\, \Delta 
\ = \ \omega - \epsilon_{d,m} 
- \langle n_{m'}\rangle \,(N-1)U +i  \Delta \;.
\label{eq:first_part_of_continued_fraction}
\end{align}
Here, the third term in the right-hand side corresponds 
to the energy shift due to the Hartree term with 
 $\langle n_{m'}\rangle = (1+r)/2$,     
the average occupation of the orbital $m'$ ($\neq m$).  
Therefore, 
the remainder part of the energy denominator 
can be regarded as the self energy correction $\Sigma_{d,m}^r(\omega)$ beyond 
the Hartree term, 
\begin{align}
& G_{m}^r(\omega) \,=\, 
\frac{1}{\omega - \epsilon_{d,m} 
- \langle n_{m'}\rangle\,(N-1)U +i \Delta  - 
\Sigma_{d,m}^r(\omega)} \;.  
\qquad \qquad 
\label{eq:selfenergy_from_continued_fraction}
\end{align}

In order to see these features of the Green's function more clearly, 
we provide some examples for first few $N$.
In the simplest case, for $N=2$, it takes the form,\cite{OguriSakano} 
\begin{align}
&
\left. G_{m}^{r}(\omega)\right|_{N=2}^{} \ =\, 
\cfrac{1}{\omega_m -  \,\frac{rU}{2} +i \Delta  - 
\cfrac{(1-r^2)\left(\frac{U}{2}\right)^2}
{\omega_m  +  \frac{rU}{2} +i 3\Delta 
}} \:.
\label{eq:A_N2}
\end{align}
For $N=3$, 
\begin{align}
&
\left. G_{m}^{r}(\omega)\right|_{N=3}^{} \ =\, 
\cfrac{1}{\omega_m -  2 \frac{rU}{2} +i \Delta  -
\cfrac{2(1-r^2)\left(\frac{U}{2}\right)^2}
{\omega_m   +i 3\Delta -    
\cfrac{2(1-r^2)\left(\frac{U}{2}\right)^2}   
{ \omega_m  + 2\frac{rU}{2} +i 5\Delta 
}}}
\;.
\label{eq:A_N3}
\end{align}
For $N=4$, 
\begin{align}
&
\left. G_{m}^{r}(\omega)\right|_{N=4}^{} \ =\, 
\cfrac{1}{\omega_m -  3 \frac{rU}{2} +i \Delta  -
\cfrac{3(1-r^2)\left(\frac{U}{2}\right)^2}
{\omega_m  -  \frac{rU}{2} +i 3\Delta -    
\cfrac{4(1-r^2)\left(\frac{U}{2}\right)^2}   
{ \omega_m  + \frac{rU}{2} +i 5\Delta - 
\cfrac{3(1-r^2)\left(\frac{U}{2}\right)^2}
{\omega_m  + 3 \frac{rU}{2} +i 7\Delta 
}}}} 
\;.
\label{eq:A_N4}
\end{align}
For $N=5$, 
\begin{align}
\left. G_{m}^{r}(\omega)\right|_{N=5}^{}  \ = \, 
\cfrac{1}{\omega_m -  4 \frac{rU}{2} +i \Delta  - 
\cfrac{4(1-r^2)\left(\frac{U}{2}\right)^2}
{\omega_m  - 2 \frac{rU}{2} +i 3\Delta -    
\cfrac{6(1-r^2)\left(\frac{U}{2}\right)^2}   
{ \omega_m  +i 5\Delta - 
\cfrac{6(1-r^2)\left(\frac{U}{2}\right)^2}
{\omega_m  + 2\frac{rU}{2} +i 7\Delta - 
\cfrac{4(1-r^2)\left(\frac{U}{2}\right)^2}
{\omega_m  + 4 \frac{rU}{2} +i 9\Delta }}}}} 
\;.
\label{eq:A_N5}
\end{align}
For $N=6$, 
\begin{align}
&
\left. G_{m}^{r}(\omega)\right|_{N=6}^{} \  =  \nonumber \\
& 
\cfrac{1}{\omega_m -  5 \frac{rU}{2} +i \Delta  - 
\cfrac{5(1-r^2)\left(\frac{U}{2}\right)^2}
{\omega_m  - 3 \frac{rU}{2} +i 3\Delta -    
\cfrac{8(1-r^2)\left(\frac{U}{2}\right)^2}   
{ \omega_m  - \frac{rU}{2} +i 5\Delta - 
\cfrac{9(1-r^2)\left(\frac{U}{2}\right)^2}
{\omega_m  + \frac{rU}{2} +i 7\Delta - 
\cfrac{8(1-r^2)\left(\frac{U}{2}\right)^2}
{\omega_m  + 3 \frac{rU}{2} +i 9\Delta -  
\cfrac{5(1-r^2)\left(\frac{U}{2}\right)^2}
{\omega_m  + 5  \frac{rU}{2} +i 11\Delta 
}}}}}} 
\;.
\label{eq:A_N6}
\end{align}
\end{widetext}
These expressions are simplified further for $r=0$, i.e.\ 
the symmetric couplings or the high-temperature limit of thermal equilibrium, 
as all the terms corresponding to the energy shift vanish.
Particularly for $N=2$, the exact self-energy becomes identical 
to the order $U^2$ results.\cite{AO2002} 
However, the similar cancellations of the higher-order terms  
in the power series of $U$ do not occur for $N>2$.
 This is because the high-order processes of  
the multiple particle-hole pair excitations occurring in different orbitals 
contribute to the self-energy for $N>2$.

%
\begin{figure}[b]
 \leavevmode
\begin{minipage}{0.9\linewidth}
\includegraphics[width=1\linewidth]{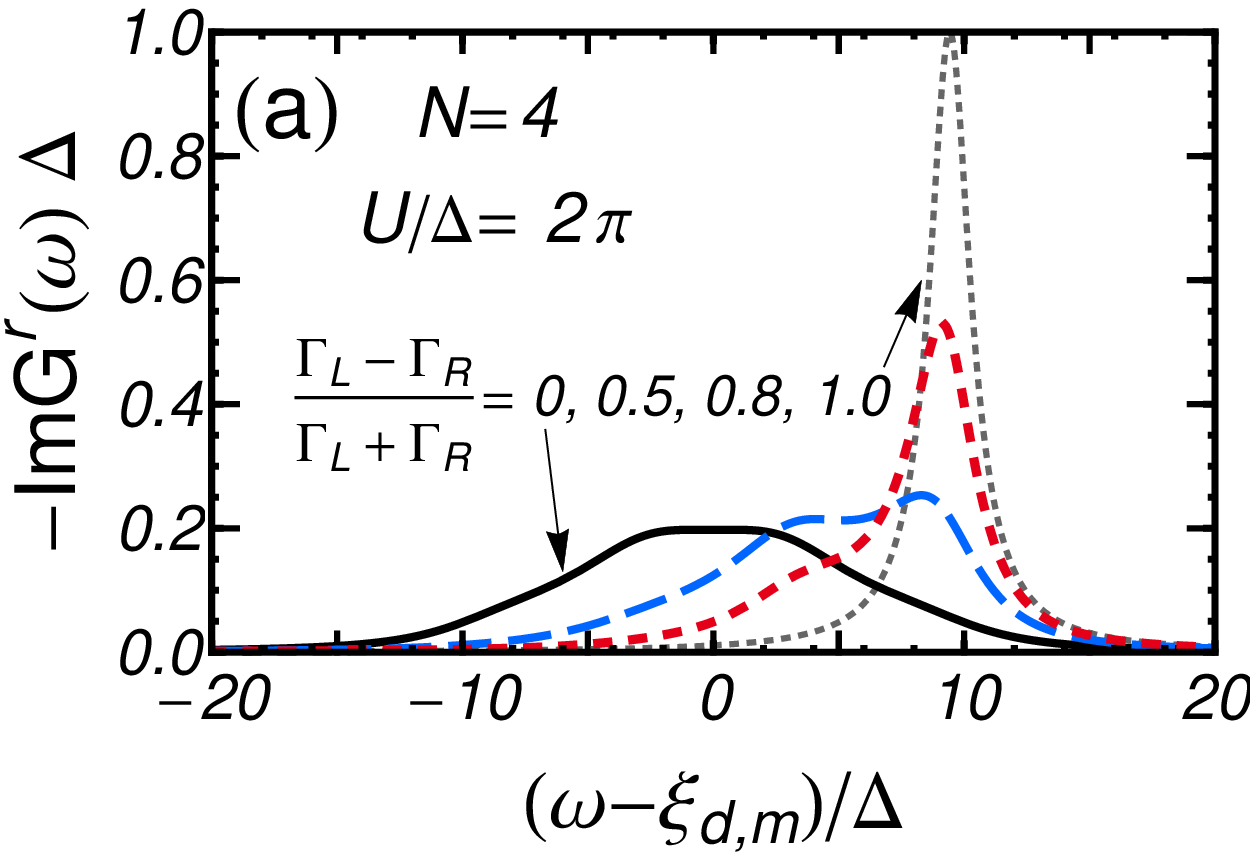}
\end{minipage}
\begin{minipage}{0.9\linewidth}
\includegraphics[width=1\linewidth]{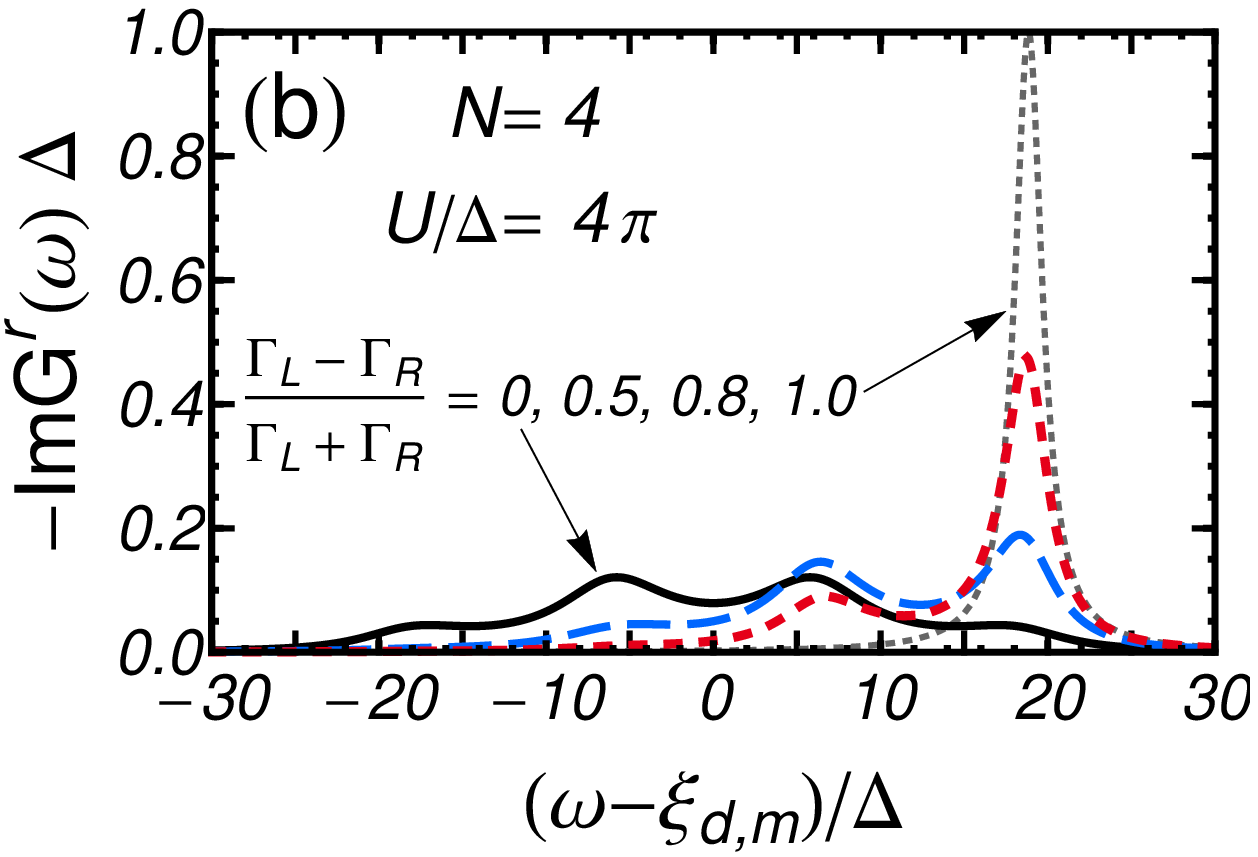}
\end{minipage}
\caption{(Color online) 
Exact high-bias spectral function for $N=4$ 
for uniform interactions (a)  
$U/(\pi \Delta) =2.0$ and (b) $4.0$, 
for different coupling asymmetries  
 $r \equiv (\Gamma_L-\Gamma_R)/(\Gamma_L+\Gamma_R)$.
}  
 \label{fig:A_N4}
\end{figure}
%
%

%
\begin{figure}[b]
 \leavevmode
\begin{minipage}{0.9\linewidth}
\includegraphics[width=1\linewidth]{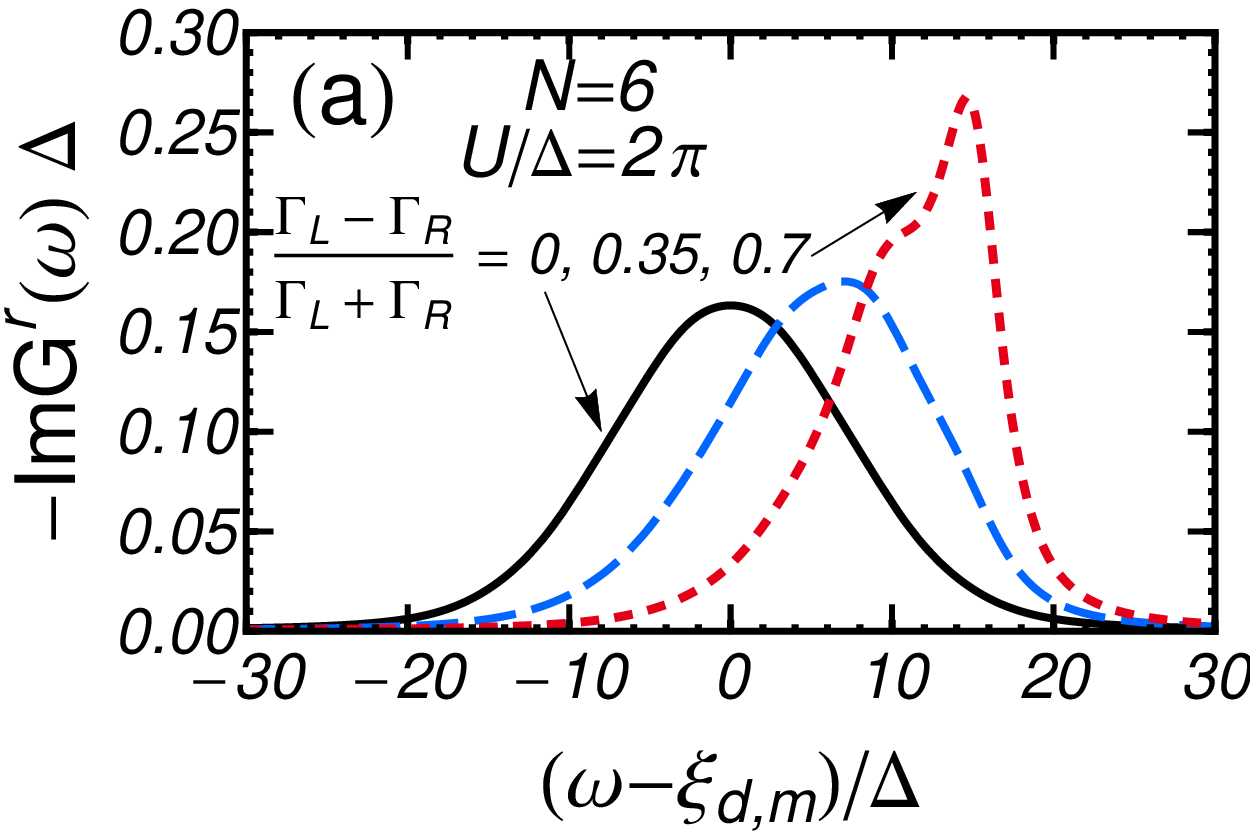}
\end{minipage}
\begin{minipage}{0.9\linewidth}
\includegraphics[width=1\linewidth]{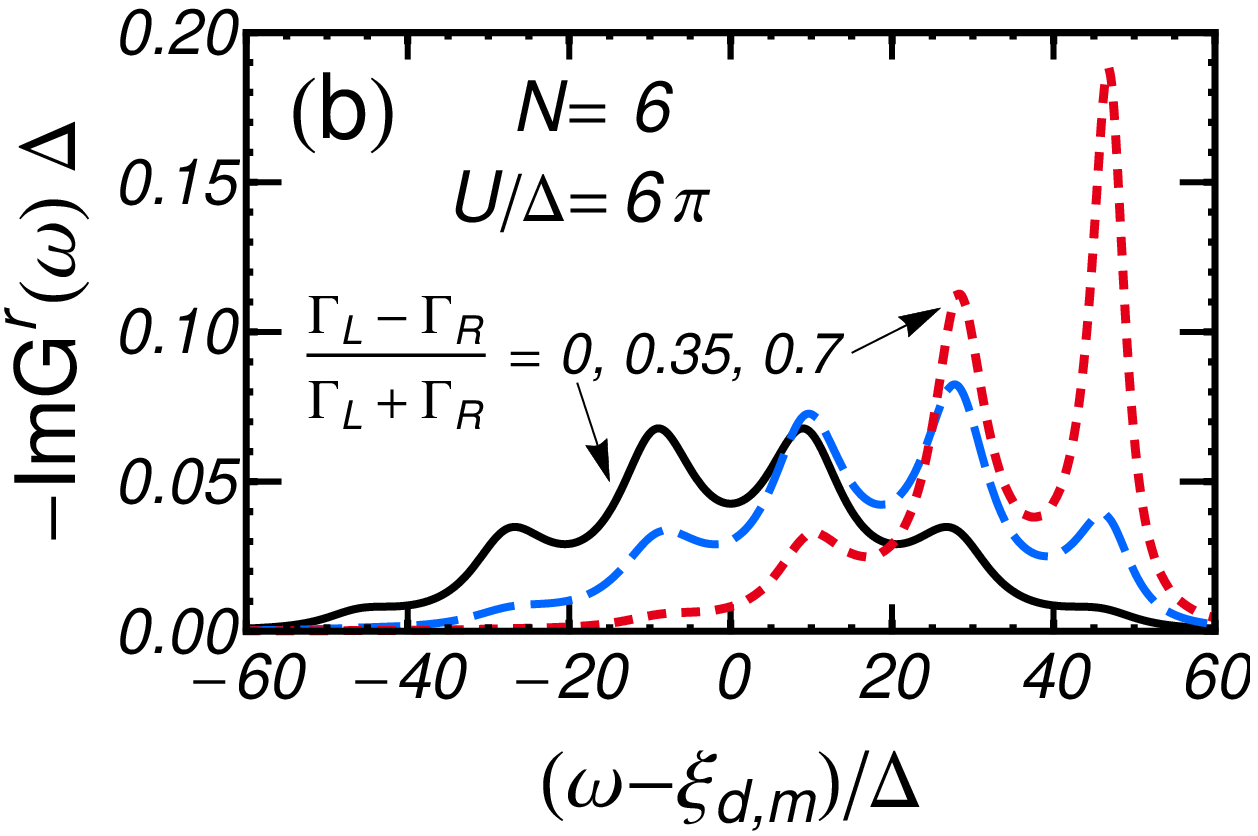}
\end{minipage}
\caption{(Color online) 
Exact high-bias spectral function for $N=6$   
for uniform interactions (a)   
$U/(\pi \Delta) =2.0$ and (b) $6.0$,  
for different coupling asymmetries  
 $r \equiv (\Gamma_L-\Gamma_R)/(\Gamma_L+\Gamma_R)$.
}  
 \label{fig:A_N6}
\end{figure}
%

\subsection{Spectral functions for the uniform interactions }
\label{subsec:U_uniform_case}

We examine further the high-bias property 
in the case of the uniform interactions. 
One of the simplest situations is 
the atomic limit where both  $\Gamma_L$ and 
$\Gamma_R$ vanish. In this case,  
the  complex eigenvalue and weight factor become 
$\mathcal{E} \to U/2$ and  $Z_{}^{(\pm)} \to 1/2$, 
respectively, as mentioned in 
Sec.\ \ref{subsec:special_limits}. 
Thus, the high-bias retarded Green's function 
given in Eq.\ \eqref{eq:retarded_G_V_inf_SUn_partial_fraction} 
simplifies 
\begin{align}
& 
\!\!\!
 G_{m}^r(\omega) 
\, \xrightarrow{\,\Gamma_{L/R}\to 0\,}   
\nonumber 
\\
 & 
 \qquad \   
\sum_{\mathcal{Q}=0}^{N-1} 
 \left( 
\begin{matrix} 
 N-1   \cr  
 \mathcal{Q} \cr
\end{matrix}          
\right)
 \frac{1}{2^{N-1}} 
\frac{1}{ \omega-\xi_{d,m}   
-\left( \mathcal{Q}-\frac{N-1}{2} \right) U } 
 \nonumber \\
& \qquad =  \frac{1}{ \omega-\xi_{d,m} - \Sigma_{d,m}^\mathrm{(ATM)}(\omega)}
.
\label{eq:retarded_G_V_inf_SUn_atomic_limit}
\end{align}
The Green's function in this limit has  
poles at $\omega = \xi_{d,m} + \bigl(\mathcal{Q}-(N-1)/2 \bigr) U$
for $\mathcal{Q}=0,1,\ldots, N-1$,
the residues of which are given by the binominal distribution. 
Each of these $N$ poles represents contributions of 
a single particle and a single hole excitations 
between the $\mathcal{Q}$-particle and $\mathcal{Q}+1$-particle states. 
This assignment of the spectrum can be verified,   
comparing 
with the equilibrium finite-temperature 
 Green's function in the atomic limit,  
given 
in Appendix  \ref{sec:atomic_finite_T}. 
The last line of 
Eq.\ \eqref{eq:retarded_G_V_inf_SUn_atomic_limit} 
defines the $T \to \infty$ 
atomic-limit self energy  $\Sigma_{d,m}^\mathrm{(ATM)}(\omega)$,  
the explicit form of which can be obtained from    
$\Sigma_{d,m}^r(\omega)$ that is defined in  
 Eqs.\ \eqref{eq:retarded_G_V_inf_SUn_continued_fraction}  
 and \eqref{eq:selfenergy_from_continued_fraction}  
 taking the limit of $\Delta \to 0$ and $r \to 0$,

The couplings to the leads, $\Gamma_L$ and $\Gamma_R$, 
make these poles resonances with finite width. 
The Hubbard I (or Hubbard II for $N>2$) approximation,
\cite{HubbardI,HubbardII} or the  decoupling approximation of 
equation of motion (EOM), gives 
the imaginary part $i \Delta$ to 
the atomic limit Green's function defined in   
Eq.\ \eqref{eq:retarded_G_V_inf_SUn_atomic_limit}. 
Specifically, in the limit of $r \to 0$ 
or high-temperature limit $T \to \infty$, 
 it appears only in the initial part of 
the continued-fraction expansion  
\begin{align}
& 
\!\!\!
 G_{m}^{r\mathrm{(EOM)}}(\omega) 
 =  \frac{1}{ \omega-\xi_{d,m} +i\Delta - \Sigma_{d,m}^\mathrm{(ATM)}(\omega)}
\;.
\end{align}
Alternatively,  
$G_{m}^{r\mathrm{(EOM)}}(\omega)$  can be expressed 
in a continued-fraction form similar to 
 Eq.\ \eqref{eq:retarded_G_V_inf_SUn_continued_fraction}, 
by replacing the coefficients $\mathcal{C}_k$ such that  
 $\mathcal{C}_1 \to 1$ for  $k=1$ and 
$\mathcal{C}_k \to 0$ for all the other $k$ ($\geq 2$).  
This indicates that the decoupling approximation of EOM 
significantly underestimates 
the relaxation effects,   especially for $N \gg 2$.

We also examine the NCA, 
which deals with the  hybridizations in a more improved way.  
Specifically, in the limit of $T \to \infty$ at equilibrium $eV=0$,  
the NCA equations for finite $U$ 
can be solved analytically as shown in Appendix \ref{sec:NCA_Tinf},  
and then the retarded Green's function takes the form
\begin{align}
& 
\!\! 
 G_{m}^{r(\mathrm{NCA})}(\omega) 
\, \xrightarrow{\,T\to \infty\,}   
\nonumber 
\\
 & 
 \ \  \sum_{\mathcal{Q}=0}^{N-1} 
 \left( 
\begin{matrix} 
 N-1   \cr  
 \mathcal{Q} \cr
\end{matrix}          
\right)
 \frac{1}{2^{N-1}} 
\frac{1}{ \omega-\xi_{d,m}   
-\left( \mathcal{Q}-\frac{N-1}{2} \right) U +iN\Delta} 
\nonumber \\
& \ = \  \frac{1}{ \omega-\xi_{d,m} 
+iN\Delta - \Sigma_{d,m}^\mathrm{(ATM)}(\omega+i N \Delta)} \;.  
\label{eq:NCA_Tinf}
\end{align}
Note that the NCA in this case takes into account all possible  
$2^N$ impurity configurations, 
from the empty to the 
fully occupied orbital states.\cite{KeiterQin}
In Eq.\ \eqref{eq:NCA_Tinf}, 
the partial-fraction representation shows that 
the spectral function   
is given by a series of the Lorentzian peaks with 
 the same width $N \Delta$. 
The last line shows that, 
 through the atomic-limit self-energy  $\Sigma_{d,m}^\mathrm{(ATM)}$ 
with the argument $\omega+i N \Delta$, 
 the constant imaginary part $i N\Delta$ also 
appears in each step of the continued fraction expansion. 
The explicit expression corresponding to 
Eq.\ \eqref{eq:retarded_G_V_inf_SUn_continued_fraction} 
can be obtained by replacing the coefficients such that 
 $\mathcal{C}_k \to N$ for all $k$ and taking $r \to 0$ as mentioned. 
Note that the exact coefficient, 
$\mathcal{C}_k =2k-1 $ given in Eq.\ \eqref{eq:A_B_C_coefficients}, 
shows that the imaginary part evolves step by step 
from $i \Delta$ to $i (2N-1) \Delta$ 
in the continued fraction expansion. 
Therefore, the constant imaginary part of $iN\Delta$ 
that the NCA gives in the limit of $T\to\infty$ corresponds 
to an average of the exact ones.

%
\begin{figure}[b]
 \leavevmode
\begin{minipage}{0.9\linewidth}
\includegraphics[width=1\linewidth]{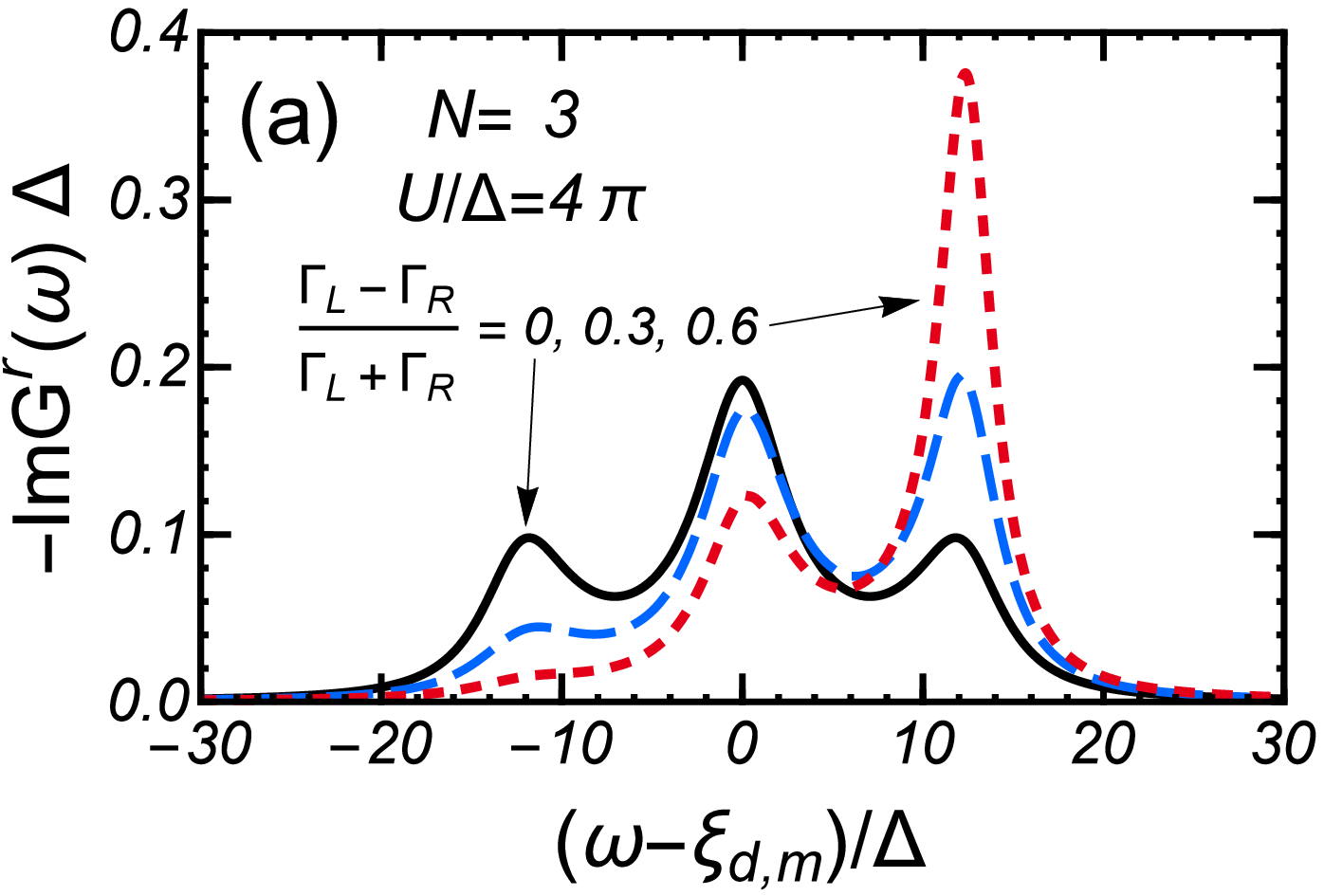}
\end{minipage}
\begin{minipage}{0.9\linewidth}
\includegraphics[width=1\linewidth]{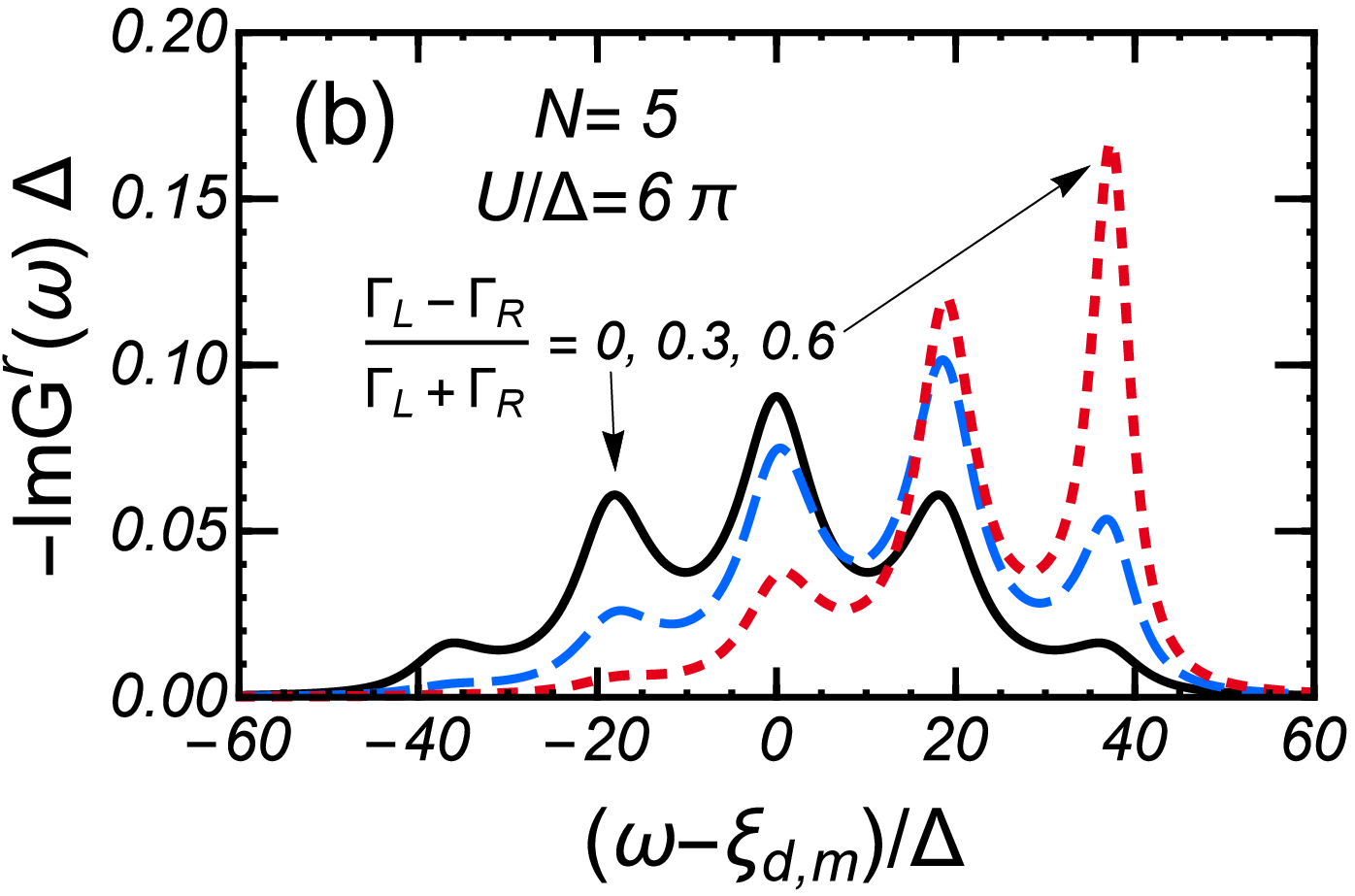}
\end{minipage}
\caption{(Color online) 
Exact high-bias spectral function 
for odd $N$ and uniform interactions:
(a) $N=3$, $U/(\pi \Delta) =4.0$,
 and (b)$N=5$, $U/(\pi \Delta) =6.0$,   
for different coupling asymmetries  
 $r \equiv (\Gamma_L-\Gamma_R)/(\Gamma_L+\Gamma_R)$.
}  
 \label{fig:A_Nodd}
\end{figure}
%

In Fig.\ \ref{fig:A_N4} the  high-bias spectral function,  
$-\mathrm{Im}\, G_{m}^{r}(\omega)$,  
for $N=4$ is plotted for some different values 
of $r$,  choosing the interactions such that 
 (a) $U/(\pi \Delta)=2.0$ and (b) $4.0$.
Four separate peaks, emerging  
at $\omega-\xi_{d,m}=\pm U/2$ and $\pm 3U/2$,  
can be recognized in Fig.\ \ref{fig:A_N4} (b) 
for symmetric coupling  $r=0$. 
As the coupling asymmetry $r$ increases, 
spectral weight moves towards a region 
around the right-end peak at $\omega-\xi_{d,m}=(N-1)U/2$
and in the limit of $r \to 1$ 
it takes the Lorentzian form with the width $\Delta$ 
which corresponds to Eq.\ \eqref{eq:limit_r_pm1}.
For negative $r$, the spectral weight 
moves in the opposite direction towards 
the left-end peak at $\omega-\xi_{d,m}=-(N-1)U/2$. 
 Note that the impurity level is 
fully occupied for $r\to 1$, or empty for $r\to -1$, 
in the case where one of the leads are disconnected.  
For weak interactions,
the level broadening 
due to the hybridizations dominates 
 and the fine structure of the spectrum is smeared   
as seen  in Fig.\ \ref{fig:A_N4} (a).

Figure \ref{fig:A_N6} shows  
another example for even $N$ ($=6$) case: 
(a) $U/(\pi \Delta)=2.0$ and (b) $6.0$. 
The six-peak spectral structure can be seen 
at $\omega-\xi_{d,m}=\pm U/2$, $\pm 3U/2$, and  $\pm 5U/2$  
for symmetric coupling $r=0$ 
 in Fig.\ \ref{fig:A_N6} (b).  
 As the coupling asymmetry $r$ increases, 
the spectral weight moves towards the higher energy region,
as mentioned in the above. 
Specifically, the results obtained at $r=0.7$ 
show a transient behavior that 
the highest two peaks share 
the most of the spectral weight.
For weak interactions, 
as seen in Fig.\ \ref{fig:A_N6} (a), 
 not all the six peaks emerge in a distinguishable way 
because of the level broadening due to the coupling to the leads.

The other examples, shown in Fig.\ \ref{fig:A_Nodd}, 
are the spectral function for odd $N$ with  
(a) $N=3$, $U/(\pi\Delta)=4.0$,
 and (b) $N=5$, $U/(\pi\Delta)=6.0$. 
In the case of odd $N$, one of the peaks appears 
at the center where $\omega-\xi_{d,m}=0$. 
The central peak corresponds to the excitations 
between the $\mathcal{Q}=(N-1)/2$ 
and $\mathcal{Q}=(N+1)/2$ particle states, 
and it is nothing  to  do with the  Kondo singlet state. 
The asymmetry in the couplings also shifts the spectral weight 
to the higher energy region as that in the even $N$ case.

\begin{figure}[b]
 \leavevmode
\begin{minipage}{0.9\linewidth}
\includegraphics[width=1\linewidth]{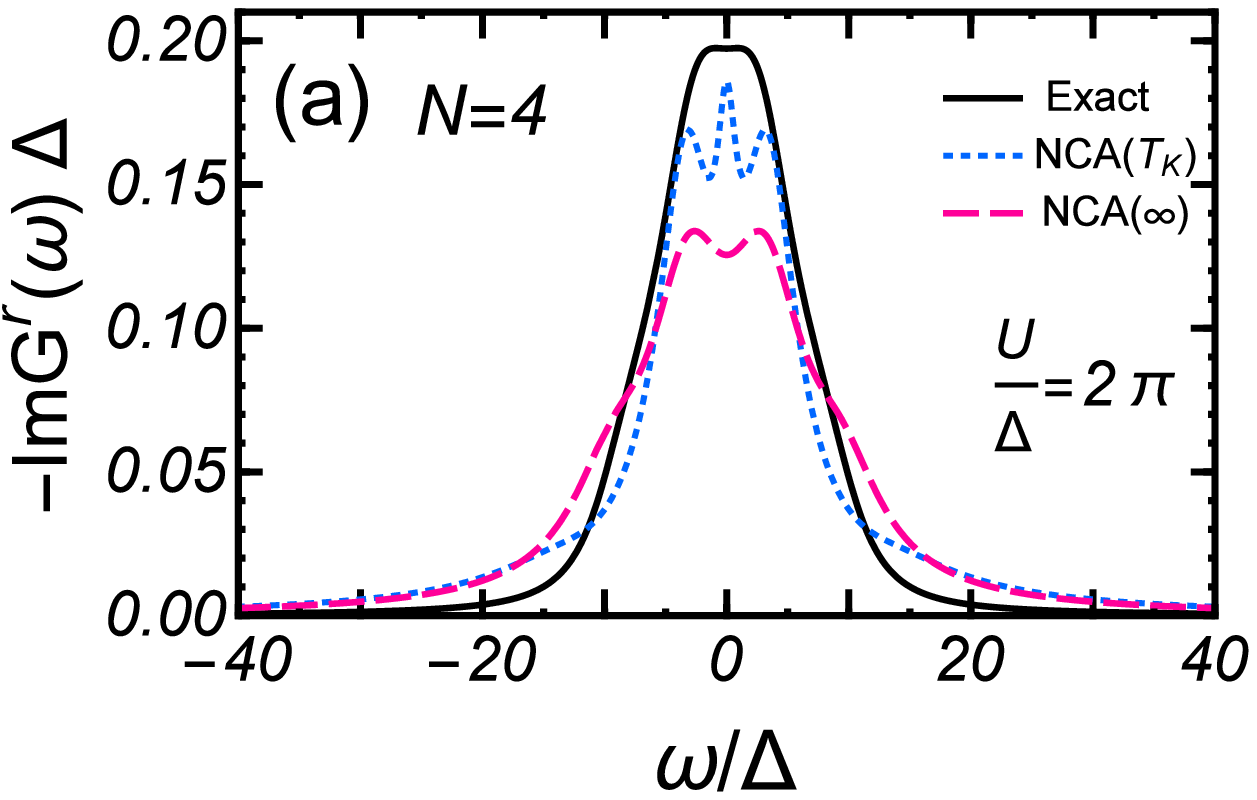}
\end{minipage}
\begin{minipage}{0.9\linewidth}
\includegraphics[width=1\linewidth]{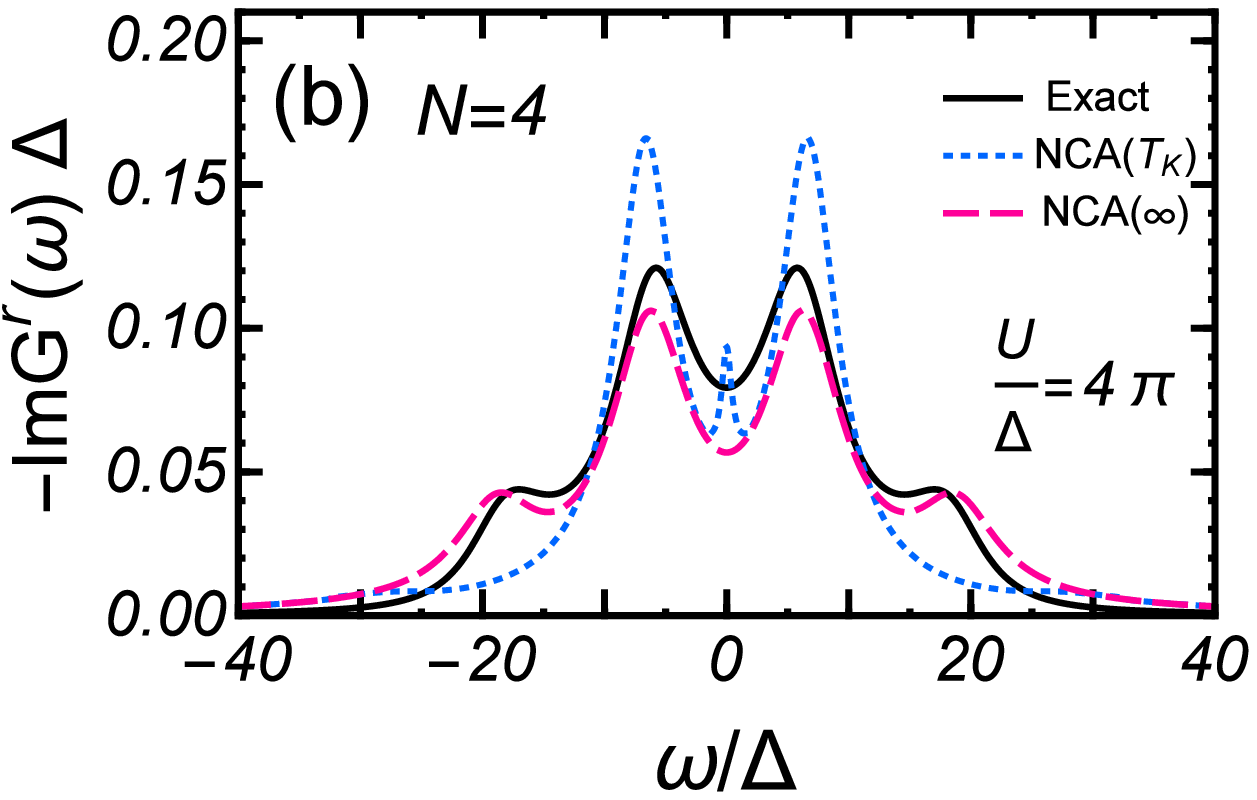}
\end{minipage}
\caption{(Color online) 
NCA and exact high-temperature  results of 
the equilibrium spectral functions in the SU(4) case for 
finite interactions, (a) $U/(\pi \Delta) = 2.0$ and (b) $4.0$,  
in the particle-hole symmetric case $\varepsilon_d=-3U/2$.   
The solid line denotes the exact $T \to \infty$ results from  
Eq.\ \eqref{eq:retarded_G_V_inf_SUn_continued_fraction}.
The NCA results are obtained  (dotted line) numerically at $T=T_K$, 
and  (dashed line) analytically at $T\to \infty$ 
from Eq.\ \eqref{eq:NCA_Tinf}.
The Kondo temperature is 
defined by $T_K = \pi z \Delta /4$,    
with $z$ the renormalization factor   
 deduced from the NRG:  (a) $z=0.52$  for $U=2.0\pi\Delta$ 
 and (b) $z=0.25$ for $U=4.0\pi\Delta$.} 
 \label{fig:A_N4_NCA}
\end{figure}

The asymptotically exact Green's function   
 can also be used as a standard  
for comparisons to check out the accuracy, 
or applicability, of theoretical calculations. 
Figure  \ref{fig:A_N4_NCA} 
compares 
the exact high-temperature results (solid line) 
and the NCA results of the spectral function 
for the SU(4) particle-hole symmetric case, 
$\varepsilon_d=-3U/2$, at equilibrium $eV=0$.
 The NCA results are obtained  
 (dotted line) at $T=T_K$ 
and (dashed line) at $T\to \infty$ from 
Eq.\ \eqref{eq:NCA_Tinf}. 
The Kondo temperature is defined by $T_K = \pi z \Delta /4$  
with $z$, the wavefunction renormalization factor that  
has been deduced from the NRG:   
 (a) $z=0.52$ for $U/(\pi \Delta)=2.0$, 
and (b) $z=0.25$ for $U/(\pi \Delta)=4.0$.
We see that the high-temperature NCA results (dashed line) 
for strong interactions, shown in Fig.\  \ref{fig:A_N4_NCA} (b),  
reasonably agree with the exact results,
specifically  at  $6.0 \lesssim |\omega|/\Delta \lesssim 12.0$ 
between the lowest and the next peaks.
However, the NCA underestimates the spectral weight 
at low frequencies  $|\omega|/\Delta  \lesssim 6.0$, 
which results in an excess accumulation of the spectral weight 
in the high-frequency region $|\omega|/\Delta  \gtrsim 18.0$  
outside the higher-energy peaks. 
Nevertheless,  for $U \gg \Delta$ 
as in the case of Fig.\  \ref{fig:A_N4_NCA} (b), 
the NCA reasonably describes  how the spectral structures  evolve 
at $T \gtrsim T_K$.
At $T=T_K$ (dotted line), 
the Kondo peak is seen  at $\omega=0$ 
with the two side peaks at $\omega=\pm U/2$ while  
the other higher-energy peaks still do not appear at $\omega=\pm 3U/2$ 
and the spectral weight spreads as a wide shoulder 
at high frequencies $|\omega| \gtrsim 3U/2$.
The higher-energy peaks evolve at high temperatures $T\gtrsim U$, and 
the NCA captures typical features of these changes. 
Similar features can also be seen in  Fig.\  \ref{fig:A_N4_NCA} (a)  
for a weak interaction. However,  
the NCA becomes less accurate for $U \lesssim \Delta$,   
where the effects of the  hybridizations dominate and 
the peak structures are smeared.


\section{Summary}
\label{sec:summary}

We have described exact high-bias properties of  
the multi-orbital Anderson impurity connected 
to two  noninteracting leads. 
In the limit of $eV \to \infty$, 
the distribution function $f_\mathrm{eff}^{(m)}(\omega)$ 
 becomes a constant independent of $\omega$, and the 
 excitations of whole energy scales equally contribute 
to the dynamics.\cite{AO2002}
Because of this highly symmetric structure 
 of the excitation processes,   
the time evolution along the Keldysh contour, 
in the high-bias limit,   
can be described by 
the effective Lagrangian of a Markovian form, 
 Eqs.\ \eqref{eq:S0_Vinf}\--\eqref{eq:SU_TFD},  
which has no long-time tail.

We have constructed the corresponding Hamiltonian formulation  
 using the non-Hermitian 
time-evolution generator $\widehat{H}_\mathrm{eff}^{}$. 
This Hamiltonian is defined with respect to the doubled Hilbert space, 
which consists of the original Fock space for the real particles  
and the counter part for the  fictitious particles 
that represent the time reversed states along the backward Keldysh contour.  
The real and fictitious particles satisfy the 
{\it boundary\/} condition in time, 
given in Eq.\ \eqref{eq:bra_boundary},
at the turnaround point $t\to\infty$ of the Keldysh contour. 
This ensures the linear dependence 
of the four components of the nonequilibrium Green's function. 
The effective Hamiltonian  $\widehat{H}_\mathrm{eff}^{}$ 
 has a highly symmetrical algebraic structure, 
Eq.\ \eqref{eq:EOM_Qtot_Vinf}, 
which can be clearly seen 
in the expression in terms of the generalized charge and current 
defined with respect to the enlarged Hilbert space.
This represents the essential symmetries 
that the excitations acquired in the high-bias limit.

We have obtained the analytic expression, \eqref{eq:retarded_G_V_inf4}, 
of the Green's function,  which  
is asymptotically exact in the $eV \to \infty$ limit. 
It shows that many-body effects on the Green's function $G_{m}^r(t)$ 
can be factorized in the time representation. 
This result holds for general 
orbital-dependent parameters; 
$\xi_{d,m}$, $\Gamma_{R,m}$, $\Gamma_{L,m}$, and $U_{m'm}$.
Furthermore, 
the continued fraction representation of $G_{m}^r(\omega)$  
 has been obtained 
for $m$-independent interactions and hybridizations.  
The explicit continued-fraction representation, given in   
Eq.\ \eqref{eq:retarded_G_V_inf_SUn_continued_fraction},
shows that 
the imaginary part emerges recursively through the relaxation of 
intermediate states with an incident particle 
accompanied by excited $k$ particle-hole pairs ($k=1,2,\ldots,N-1$),  
which give the damping rate of $(2k-1)\Delta$.

The corresponding spectral function has  
$N$ separate peaks at $\omega-\xi_{d,m}=-(N-1)U/2, \ldots (N-1)U/2$ 
for symmetric coupling $\Gamma_L = \Gamma_R$ with 
strong interactions $U \gg \Delta$. 
The coupling asymmetry $\Gamma_L \neq \Gamma_R$ varies 
the average impurity occupation,  
and shifts the spectral weight towards high-energy region. 
We have also examined the temperature 
dependence of the spectral weight using the NCA,  
which can be analytically solved for $T \to \infty$.
The results demonstrate a typical feature:  
among the $N$ separate peaks seen in the limit of $T \to \infty$ 
the ones corresponding to the highest energy excitations 
disappear as temperature decreases, 
and for large $N$ the next-highest ones will also disappear 
as $T$ decreases further.
Our results can also be used as a standard to check  
theoretical approaches to 
out-of-equilibrium  
quantum impurities at high bias voltages.

\begin{acknowledgments}

This work was supported by JSPS KAKENHI Grant Numbers 
 26400319, 24540316, 26220711, and 25800174.


\end{acknowledgments}

\appendix

\section{Feynman rule for the Hartree term}
\label{sec:Hartree}

\begin{figure}[t]
 \leavevmode
\begin{minipage}{0.5\linewidth}
%
\includegraphics[trim = 50 0 50 0, width=1\linewidth]{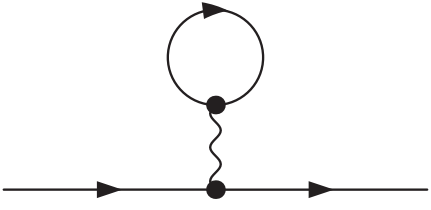}
\end{minipage}
\caption{
Feynman diagram for the Hartree term.}
 \label{fig:Hartree}
\end{figure}

There is a slight difference between  
the Feynman rules for the Keldysh Green's function 
 $G^{\mu\nu}_{m}$ and 
  those for the Green's function  
 $\mathcal{G}^{\mu\nu}_{m}$ defined with respect to 
the doubled Hilbert space. 
It emerges for the  $+$ component of 
the Hartree-type self-energy $\Sigma^{++}_{m}$,
which corresponds to the tadpole diagram shown in Fig.\ \ref{fig:Hartree}.
As the arguments $t$ and $t'$ for the inner Green's function 
along the loop are equal, 
the limit is required to be taken carefully  
 such that $G^{++}_{m}(t+ 0^+,t)$ 
in the Keldysh approach  
 whereas the opposite limit is required 
for $\mathcal{G}^{++}_{m}(t,t+ 0^+)$ 
in the thermal-field-theoretical approach.  
  This is caused by the difference 
in the direction of the time-ordering for 
the operators belonging to the $+$ branch. 
Thus, for the $-$ component of
the Hartree-type self-energy $\Sigma^{--}_{m}$,
the same limit $t'\to t+0^+$ is taken  
for both the Keldysh 
and the thermal-field-theoretical Green's functions.

The effective Hamiltonian $\widehat{H}_\mathrm{eff}^{}$, 
defined in Eqs.\ \eqref{eq:H_0_Vinf} and \eqref{eq:H_U_Vinf}, 
includes the $U$-dependent terms such that  
\begin{align}
& 
\frac{1}{2} \sum_{m\neq m'} U_{mm'}\,Q_m^{} \! 
\,+ \,\widehat{H}_\mathrm{eff}^{(U)}  
\nonumber \\
& =   \  
\frac{1}{2} \sum_{m\neq m'} U_{mm'}
\left( n_{-,m} \,n_{-,m'}  -   
n_{+,m}\,n_{+,m'} \right)
+ \widehat{H}_\mathrm{eff}^{(\mathrm{CT})}  .
\end{align}
The last term includes only the number operators for the $+$ branch, 
\begin{align}
 \widehat{H}_\mathrm{eff}^{(\mathrm{CT})}  
\equiv  \  
\sum_{m=1}^N \sum_{m'(\neq m)} U_{mm'}
\left( n_{+,m}  -\frac{1}{2} \right) 
\;,
\end{align}
and can be regarded a counter term for the particles in the $+$ branch.
This term compensates the difference that 
 arises in the $+$ component of the Hartree energy shift, 
mentioned above.

\section{Atomic limit in thermal equilibrium}
\label{sec:atomic_finite_T}

The atomic-limit Green's function  in equilibrium 
takes the following 
form at finite temperatures,
\begin{align}
 &
 G_{m}^\mathrm{(ATM)}(\omega)
  \,=  \,   
 \sum_{\mathcal{Q}=0}^{N-1} 
  \left( 
 \begin{matrix} 
  N-1   \cr  
  \mathcal{Q} \cr
 \end{matrix}          
 \right) 
 \frac{1}{\Xi}\,
 \frac{e^{-\beta E_{\mathcal{Q}+1}} + e^{-\beta E_{\mathcal{Q}}}}
 { \omega - 
 \left(E_{\mathcal{Q}+1}- E_{\mathcal{Q}}\right)
 } 
 \;, 
 \label{eq:retarded_G_V_inf_SUn_atomic_finite_T}
\\
&\Xi =  
 \sum_{\mathcal{Q}=0}^{N} 
 \left( 
 \begin{matrix} 
  N   \cr  
  \mathcal{Q} \cr
 \end{matrix}          
 \right) 
 e^{-\beta E_{\mathcal{Q}}} , 
\qquad 
 E_{\mathcal{Q}}  =    
 \mathcal{Q}\, \varepsilon_{d} 
 + \frac{U}{2}  \,\mathcal{Q}  \left( \mathcal{Q} -1 \right) 
\end{align}
in the SU($N$) case  
where quantized level, $\varepsilon_{d}$, 
has $N$-fold degeneracy.\cite{HubbardII}
This function has the poles  at  
$ \omega= \varepsilon_{d} + \,\mathcal{Q}U$ for  $\omega$ for $\mathcal{Q}=0,1,2,\ldots,N-1$. 
 Equation \ \eqref{eq:retarded_G_V_inf_SUn_atomic_finite_T}
coincides with 
Eq.\ \eqref{eq:retarded_G_V_inf_SUn_atomic_limit} 
in the $T\to \infty$ limit.

\section{Noncrossing approximation}
\label{sec:NCA_Tinf}

The closed system of equations of the NCA can  be 
analytically solved 
in the high-temperature limit $T \to \infty$ at equilibrium $eV=0$, 
to yield the expression,  given in Eq.\ \eqref{eq:NCA_Tinf}. 
In this appendix, we provide the outline of derivation. 

\subsection{Basic equations of the NCA}

The NCA is a self-consistent
perturbation theory, 
which collects a specific series of expansions in the hybridization.
\cite{Bickers,KeiterQin,Kroha,OtsukiKuramoto} 
This method is known to give physically reasonable result 
at energy scales near the Kondo temperature.
To work in this approximation,
we rewrite the Hamiltonian, given by 
Eqs.\ \eqref{Hami_seri_part} and \eqref{eq:mia-model}, in the form,
\begin{eqnarray}
{\cal H} &=& {\cal H}_{\rm band}^{} + {\cal H}_{\rm dot}^{} 
+ {\cal H}_{\rm hyb}^{} \\
{\cal H}_{\rm band}^{} &=& \sum_{m=1}^{N} 
\int_{-D}^{D} d \epsilon\, 
\epsilon\, c_{\epsilon m}^{\dagger} c_{\epsilon m}^{} \;, \\
{\cal H}_{\rm dot}^{} &=& 
\sum_{n=1}^{2^N} E_n^{} \left| n \right\rangle \!\! \left\langle n \right| \;, 
\\
{\cal H}_{\rm hyb}^{} &=& \sum_{m=1}^{N} \sum_{n,n'}^{2^N}
 \left( v_{m}^{} M_{n,n'}^{m} \left| n \right\rangle \!\! 
\left\langle n' \right|  
\psi_{m}^{} + \mbox{H.c.} \right) .
\end{eqnarray}
Here,  $E_n$ and $\left| n \right\rangle$ are
the $2^N$ many-body eigenvalues and corresponding eigenstates 
 of ${\cal H}_{\rm dot}^{}$  
that include the interactions between electrons in the dot.
The matrix element 
$M_{n,n'}^{m} \equiv \left\langle n \right| 
d_m^{\dagger} \left| n'\right\rangle$ 
is defined between these many-body eigenstates.
Note that we consider an equilibrium situation, 
and therefore only a linear combination of the the conduction bands 
which couples to the dot  
$\psi_{m}^{} \equiv 
(v_{L,m}^{}\psi_{L,m}^{}+v_{R,m}^{}\psi_{R,m}^{})/v_{m}^{}$ 
with $v_{m}^{}= \sqrt{v_{R,m}^{2}+v_{L,m}^{2}}$ 
are explicitly shown in the above Hamiltonian.

The NCA for finite interactions can be described by 
the coupled equations for the retarded 
resolvents and the self-energies,
\begin{eqnarray}
R_n(\omega) &=&  \frac{1}{\omega - E_n - \Sigma_n^\mathrm{(NCA)} (\omega)} \;, 
\\
\Sigma_n^\mathrm{(NCA)} (\omega) &=&  
\sum_{n'=1}^{2^N} \sum_{m=1}^{N} 
\frac{\Delta_{m}^{}}{\pi} 
\left[ \left( M_{n,n'}^{m} \right)^2 
+\left( M_{n',n}^{m} \right)^2 \right]
\nonumber \\
&& \qquad \quad \times \int_{-D}^{D} d\epsilon\, 
R_{n'}(\omega + \epsilon) f (\epsilon) \;.
\label{eq:rslvnt-slfenrgy}
\end{eqnarray}

The local density of states at the dot site is given by
\begin{eqnarray}
\!\!\!\!\! 
\rho_{dm}^{} (\omega) &=&  - \frac{1}{\pi} 
{\rm Im} \, G_{m}^r (\omega) \nonumber \\
&=& \frac{1}{Z_{\rm total}/Z_{\rm band}} 
\sum_{n,n'}^{2^N} \left( M_{n,n'}^{m} \right)^2
\int_{-D}^{D} d\epsilon \, e^{-\beta\epsilon} \nonumber \\
&&  \times 
\Bigl[ \rho_n(\epsilon) \rho_{n'}(\epsilon + \omega) 
+\rho_n(\epsilon) \rho_{n'} (\epsilon - \omega)  \Bigr],
\label{eq:NCA_spectral_generic}
\end{eqnarray}
with a partition function
\begin{eqnarray}
\frac{Z_{\rm total}}{Z_{\rm band}} = \sum_{n=1}^{2^N} \int_{-D}^{D} 
d\epsilon\, e^{- \beta \epsilon} \rho_n (\epsilon) \;, 
\end{eqnarray}
and the spectral function for the resolvent 
\begin{eqnarray}
\rho_n (\omega) = -\frac{1}{\pi}\, {\rm Im}\, R_n (\omega) \;.
\end{eqnarray}

\subsection{High-temperature limit}
In the high-temperature limit,
the Fermi distribution function in Eq.\ (\ref{eq:rslvnt-slfenrgy})
is replaced by a constant $1/2$,
and the integration can be readily executed to give an $\omega$ independent 
constant.
Then the NCA equation can be solved and the resolvent is given by a 
Breit-Wigner form
\begin{eqnarray}
R_n(\omega) = \frac{1}{\omega - E_n + i \Gamma_n} \;, 
\end{eqnarray}
with
\begin{eqnarray}
\Gamma_n = \sum_{n'}^{2^N} \sum_m^{N} \frac{\Delta_m}{2}  \left[ \left( M_{n,n'}^{m} \right)^2 +\left( M_{n',n}^{m} \right)^2 \right] \;.
\end{eqnarray}
Substituting these forms into Eq.\ \eqref{eq:NCA_spectral_generic}, 
the asymptotic form of the local density of state 
in the limit of $T \to \infty$, can be expressed in a sum of 
the Lorentzian peaks,  
\begin{eqnarray}
\!\!\!\!\!\!\!\! 
\rho_{dm}^{} (\omega)
&=& \frac{1}{2^{N-1}} \sum_{n,n'}^{2^N} \left( M_{n,n'}^{m} \right)^2
\frac{\Gamma_n + \Gamma_{n'}}{\pi}  \nonumber \\
&&  \times
\frac{1}{\bigl[\,\omega - (E_{n'}^{} - E_n^{})\,\bigr]^2 + \left( \Gamma_n+\Gamma_{n'} \right)^2 } .
\label{eq:NCA_spectral_Tinf_app}
\end{eqnarray}

Particularly, for the $m$ independent interactions and hybridizations, 
 $U_{mm'} \equiv U$ and $\Delta_m \equiv \Delta$,   
the asymptotic expressions can be more simplified 
because $\Gamma_n$ in this case is explicitly in the form 
\begin{eqnarray}
\Gamma_n = \frac{N\Delta}{2}.
\end{eqnarray}
Then, Eq.\ \eqref{eq:NCA_spectral_Tinf_app}  
corresponds to the NCA Green's function 
given in Eq.\ \eqref{eq:NCA_Tinf}.



\begin{thebibliography}{99}
\bibitem{TFD} 
We are using a standard notation of the thermal field theory 
and the Liouville-Fock space approach for the initial and final states, i.e.\ 
$\langle \! \langle I |\! |$ and  $|\! | \rho \rangle \! \rangle$.
\cite{EzaAriHas,EspositeHarolaMukamel,EzaAriHas} 
With these approaches, 
a mixed-state average defined in the original Hilbert space 
can be reformulated as a pure-state average with respect to 
these two states in the enlarged Hilbert space. 
The statistical density weight can be 
described through $|\! | \rho \rangle \! \rangle$
while $\langle \! \langle I |\! |$
plays a role of the projection operator onto the physical subspace.
\end{thebibliography}
\end{document}